\documentclass[11pt]{article}
\usepackage{amsmath}
\usepackage{graphicx,psfrag,epsf}
\usepackage{enumerate}
\usepackage{psfrag,epsf}
\usepackage{url} 
\usepackage{amsmath,amssymb,amsfonts,amsthm,mathtools,mathrsfs}

\usepackage{graphicx}
\usepackage{caption}
\usepackage{subcaption}
\usepackage{booktabs}
\usepackage{makecell}
\usepackage{booktabs}
\usepackage[table]{xcolor}
\usepackage{siunitx}
\usepackage{bm,bbm}
\usepackage{color}

\usepackage{algorithm,algpseudocode}
\usepackage[symbol]{footmisc}
\usepackage{scalefnt}
\usepackage{authblk} 
\usepackage{multirow,centernot}
\usepackage[colorlinks=true, citecolor=blue, urlcolor=blue]{hyperref}
\usepackage{booktabs}
\usepackage[table]{xcolor}

\usepackage{epstopdf} 

\graphicspath{{figures/}}
\DeclareGraphicsExtensions{.pdf,.eps,.png,.jpg}

\usepackage{natbib}
\usepackage{dsfont}
\usepackage[title]{appendix}
\usepackage{booktabs,longtable}

\usepackage{newtxtext,newtxmath} 
\usepackage{mhchem} 

\input cyracc.def

\usepackage[left=1in,top=1.1in,right=0.5in,bottom=1in]{geometry}

\theoremstyle{definition}

\newtheorem{theorem}{Theorem}[section]

\newtheorem{remark}[theorem]{Remark}

\makeatletter
\def\@seccntformat#1{\@ifundefined{#1@cntformat}%
	{\csname the#1\endcsname\quad}
	{\csname #1@cntformat\endcsname}
}
\makeatother

\newif\ifShowComments
\ShowCommentstrue
\def\strutdepth{\dp\strutbox}
\def\druk#1{\strut\vadjust{\kern-\strutdepth
        {\vtop to \strutdepth{%
                \baselineskip\strutdepth\vss
                        \llap{\hbox{#1}\quad}\null}}}}

\title{\bf
Closed-form parameter estimation for the bivariate gamma distribution: New approaches

}

\author[1]{Roberto Vila \thanks{Corresponding author: rovig161@gmail.com}}
\author[1,2]{Helton Saulo}
\affil[1]{Department of Statistics, University of
	Brasilia, Brasilia, Brazil}
\affil[2]{Department of Economics, Federal University of Pelotas, Pelotas, Brazil}
\begin{document}
	\maketitle 	
	\begin{abstract}
We propose new closed-form estimators for the parameters of McKay’s bivariate gamma distribution by exploiting monotone transformations of the likelihood equations. As a special case, our framework recovers the estimators recently introduced by \cite{Zhao2022} [Zhao, J., Jang, Y.-H., and Kim, H. (2022). Closed-form and bias-corrected estimators for the bivariate gamma
distribution. Journal of Multivariate Analysis, 191:105009]. Theoretical properties, including strong consistency and asymptotic normality, are established. We further introduce a second family of closed-form estimators that is explicitly built from the stochastic relationship between gamma random variables. Our second approach encompasses the estimators of \cite{Nawa2023} [Nawa, V. M. and Nadarajah, S. (2023). New closed form estimators for a bivariate gamma distribution. Statistics,
57(1):150–160]. Monte Carlo experiments are conducted to assess finite-sample performance, showing that the new estimators perform comparably to maximum likelihood estimators while avoiding iterative optimization, and improve upon the existing closed-form approach by \cite{Zhao2022} and \cite{Nawa2023}. A real hydrological data set is analyzed to illustrate the proposed approaches.
\end{abstract}

\smallskip
	\noindent
	{\small {\bfseries Keywords.} {McKay's bivariate gamma distribution $\cdot$ Closed-form estimators $\cdot$Maximum likelihood method $\cdot$ Monte Carlo simulation.}}
	\\
	{\small{\bfseries Mathematics Subject Classification (2010).} {MSC 60E05 $\cdot$ MSC 62Exx $\cdot$ MSC 62Fxx.}}
	
{
	\hypersetup{linkcolor=black}
	\tableofcontents
}

\section{Introduction}
	\noindent
The bivariate gamma distribution introduced by \cite{McKay1934} has played an important role
in hydrological applications; see, for example, \cite{Yue2001}. Along these lines, \cite{Gupta2006} derived the exact distributions of $R = X+Y$, $U = XY$, and $W = X/(X+Y)$, together with their moment properties,
when $(X,Y)^\top$ follows McKay’s bivariate gamma distribution,
motivated primarily by applications in hydrology. \cite{Gupta2006b} derived the  exact distributions for $P = XY$ and their associated moment properties when $X$ and $Y$ are correlated gamma random variables arising from bivariate gamma distributions.

Traditionally, the parameters of the bivariate gamma distribution are estimated by maximum likelihood (ML). While asymptotically efficient, ML requires iterative optimization, which can be computationally intensive and prone to numerical instabilities in small samples. Recently, \cite{Zhao2022} proposed moment-type estimators by applying the likelihood function of a transform of
the bivariate gamma variable. However, these estimators may suffer from non-negligible bias in small samples and bias-corrected estimators have also proposed to deal with it. Motivated by \cite{Zhao2022}, \cite{Nawa2023} proposed simpler closed-form estimators for the bivariate gamma distribution, with smaller asymptotic variances and covariances, and improved performance in real-data applications.

The purpose of this paper is to advance closed-form estimation methods for the McKay bivariate gamma distribution. We first build on a general framework for generating estimators via monotone transformations of the likelihood equations \citep{Vila2025}, leading to a broad class of closed-form estimators for the distribution’s parameters. This approach not only generalizes the estimators proposed by \cite{Zhao2022}, but also introduces new ones with superior finite-sample performance. We establish that these estimators are strongly consistent and asymptotically normal. As a second contribution, we develop an alternative class of closed-form estimators by leveraging the stochastic relationships implied by the McKay construction. This second approach subsumes the estimators of \cite{Nawa2023} as special cases, further broadening the scope of closed-form methods available for this distribution.

The performance of the proposed estimators is evaluated through an extensive Monte Carlo study. The results indicate that the new estimators achieve accuracy comparable to ML while avoiding the burden of iterative numerical routines, and that they outperform the methods of \cite{Zhao2022} and \cite{Nawa2023}. These findings suggest that the proposed classes of estimators provide simple and efficient alternatives for inference in the bivariate gamma model.

The remainder of the paper is structured as follows. Section~\ref{sec:power} introduces a power-transformed extension of the bivariate gamma distribution, which serves as a foundation for deriving the proposed estimators. In Section~\ref{The New Estimators}, we present the first class of closed-form estimators, constructed using monotone transformations of the likelihood equations. Section~\ref{largeprop} establishes the large-sample properties of these estimators, including strong consistency and asymptotic normality. In Section~\ref{The New Estimators-1}, we propose a second class of estimators, developed by exploiting stochastic relationships inherent to the McKay construction. Section~\ref{sec:simulation} reports results from extensive Monte Carlo simulations that assess the finite-sample performance of the proposed and existing methods. In Section~\ref{sec:application}, we apply the proposed estimators to Los Angeles annual rainfall data to demonstrate the usefulness of the estimation methods, and
finally in Section \ref{sec:conclusion}, we provide some concluding remarks.

\section{A power-transformed extension of the bivariate gamma distribution}\label{sec:power}


A two-dimensional random vector $(X,Y)^\top$ has the McKay’s bivariate gamma distribution \citep{McKay1934} if its joint probability density function (PDF) is given by
\begin{align}\label{pdf-1}
f(x,y;\boldsymbol{\theta})
=
{\gamma^{\alpha+\beta}\over\Gamma(\alpha)\Gamma(\beta)}\,
x^{\alpha-1}(y-x)^{\beta-1}
\exp(-\gamma y),
\quad 
y>x>0,
\end{align}
where $\boldsymbol{\theta}=(\alpha,\beta,\gamma)^\top$ is the parameter vector and $\alpha,\beta,\gamma>0$.
It is not difficult to verify that a  random vector $(X,Y)^\top$ with PDF \eqref{pdf-1} has the following stochastic representation: 
\begin{align}\label{rep-stoch}
	X=X_1,
	\quad
	Y=X_1+X_2,
\end{align}
where $X_1\sim{\rm Gamma}(\alpha,\gamma)$ and $X_2\sim{\rm Gamma}(\beta,\gamma)$ (with $\alpha,\beta$ shapes and
$\gamma$ rate) are independent. Under these conditions, it is clear that $Y=X_1+X_2\sim{\rm Gamma}(\alpha+\beta,\gamma)$.

\bigskip 


We now extend the bivariate gamma distribution via a two-dimensional power transformation.
Let $U=[g_1^{-1}(X)]^{1/p}$ and $V=[g_2^{-1}(Y)]^{1/q}$ be two monotonic transformations, with $p>0$ and $q>0$, as considered in \cite{Vila2025}.
Here, each function $g_i: D\subset (0,\infty)\to \text{supp}(X,Y)$, for $i=1,2$, is assumed to be  twice-differentiable and monotonic, with inverse  $g^{-1}_i$.
The support of the random vector $(X,Y)^\top$, denoted by $\text{supp}(X,Y)=\{(x,y)\in(0,\infty)^2:y>x>0\}$, 
corresponds to the domain of the joint PDF given \eqref{pdf-1}. It is clear that the joint PDF of $(U,V)^\top$ is written as
\begin{eqnarray}\label{dist-gen-exp}
f(u,v;\boldsymbol{\theta},p,q)
=
{pq\gamma^{\alpha+\beta}\over\Gamma(\alpha)\Gamma(\beta)}\,
u^{p-1}v^{q-1} \vert g_1'(u^p)\vert \vert g_2'(v^q)\vert
g_1^{\alpha-1}(u^p)
[g_2(v^q)-g_1(u^p)]^{\beta-1}
\exp\{-\gamma g_2(v^q)\},
\end{eqnarray}
where $(u,v)^\top\in D^2$ and $\boldsymbol{\theta}=(\alpha,\beta,\gamma)^\top$.
From \eqref{rep-stoch}, the random vector $(U,V)^\top$ admits the stochastic representation 
$U=[g_1^{-1}(X_1)]^{1/p}$, $V=[g_2^{-1}(X_1+X_2)]^{1/q}$,
where $X_1\sim{\rm Gamma}(\alpha,\gamma)$ and $X_2\sim{\rm Gamma}(\beta,\gamma)$ are independent.

\section{New closed-form estimators: First approach}\label{The New Estimators}

Let $\{(U_i,V_i)^{\top} : i = 1,\ldots , n\}$ be a bivariate random sample of size $n$ from  $(U,V)^{\top}$ having joint PDF \eqref{dist-gen-exp}.
%
The (random) likelihood function $L\equiv L(\boldsymbol{\theta},p,q)$ for $(\boldsymbol{\theta},p,q)^\top$ is written as
\begin{eqnarray*}
L   
=
\left[
{pq\gamma^{\alpha+\beta}\over\Gamma(\alpha)\Gamma(\beta)}
\right]^n
\prod_{i=1}^{n}
\left\{
U^{p-1}_i V^{q-1}_i 
\vert g_1'(U_i^p)\vert \vert g_2'(V_i^q)\vert
g_1^{\alpha-1}(U_i^p) [g_2(V^q_i)-g_1(U^p_i)]^{\beta-1}
\right\}
\exp\left(-\gamma \sum_{i=1}^{n}g_2(V^q_i)\right).  
\end{eqnarray*}
Consequently, the (random) log-likelihood function $l(\boldsymbol{\theta},p,q)\equiv\log(L)$ for $(\boldsymbol{\theta},p,q)^\top$ is written as
\begin{align*}
l(\boldsymbol{\theta},p,q)   
&=
n\left[
\log(p)+\log(q)
+
(\alpha+\beta)\log(\gamma)
-
\log(\Gamma(\alpha))
-
\log(\Gamma(\beta))
\right]
\\[0,2cm]
&
+
(p-1)
\sum_{i=1}^{n}
\log(U_i)
+
(q-1)
\sum_{i=1}^{n}
\log(V_i)
+
\sum_{i=1}^{n}
\log(\vert g_1'(U_i^p)\vert)
+
\sum_{i=1}^{n}
\log(\vert g_2'(V_i^q)\vert)
\\[0,2cm]
&
+
(\alpha-1)
\sum_{i=1}^{n}
\log(g_1(U_i^p))
+
(\beta-1)
\sum_{i=1}^{n}
\log(g_2(V_i^q)-g_1(U_i^p))
-
\gamma
\sum_{i=1}^{n} g_2(V^q_i).
\end{align*}

Through elementary calculus, one finds that the components of the (random) score vector take the form:
\begin{align*}
 {\partial l(\boldsymbol{\theta},p,q) \over\partial \alpha}
 &=
 n\left[
\log(\gamma)-\psi^{(0)}(\alpha)
+
{1\over n}
\sum_{i=1}^{n}
\log(g_1(U_i^p))
\right],  
 \\[0,2cm]
  {\partial l(\boldsymbol{\theta},p,q) \over\partial \beta}
 &=
 n
 \left[
 \log(\gamma)-\psi^{(0)}(\beta)
 +
{1\over n}
\sum_{i=1}^{n}
\log(g_2(V_i^q)-g_1(U_i^p))
\right],  
 \\[0,2cm]
  {\partial l(\boldsymbol{\theta},p,q) \over\partial \gamma}
 &=
 n\left[
  {\alpha+\beta\over\gamma} 
  -
{1\over n}
\sum_{i=1}^{n} g_2(V^q_i)
  \right],
 \\[0,2cm]
  {\partial  l(\boldsymbol{\theta},p,q) \over\partial p}
 &=
 n
 \Bigg[
 {1\over p}
 +
 {1\over n}
\sum_{i=1}^{n}
\log(U_i)
+
{1\over n}
\sum_{i=1}^{n}
{g_1''(U_i^p) U_i^p\log(U_i)\over g_1'(U_i^p)} 
\\[0,2cm]
&
+
(\alpha-1) \,
{1\over n}
\sum_{i=1}^{n}
{g_1'(U_i^p)U_i^p\log(U_i)\over g_1(U_i^p)} 
-
(\beta-1) \,
{1\over n}
\sum_{i=1}^{n}
{g_1'(U_i^p)U_i^p\log(U_i)\over g_2(V_i^q)-g_1(U_i^p)} 
\Bigg], 
 \\[0,2cm]
{\partial  l(\boldsymbol{\theta},p,q) \over\partial q}
&=
n
\Bigg[
{1\over q}
+
 {1\over n}
\sum_{i=1}^{n}
\log(V_i)
+
{1\over n}
\sum_{i=1}^{n}
{g_2''(V_i^q)V_i^q\log(V_i)\over g_2'(V_i^q)} 
\\[0,2cm]
&+
(\beta-1) \,
{1\over n}
\sum_{i=1}^{n}
{g_2'(V_i^q)V_i^q\log(V_i)\over g_2(V_i^q)-g_1(U_i^p)} 
-
\gamma \,
{1\over n}
\sum_{i=1}^{n}
{g_2'(V_i^q)V_i^q\log(V_i)}
\Bigg],
\end{align*}
where $\psi^{(0)}(x)=\partial \log(\Gamma(x))/\partial x$ is the polygamma function of order $0$.
Since $g_1(U^p_i)=X_i$, $U_i^p=g_1^{-1}(X_i)$, $g_2(V^q_i)=Y_i$ and $V_i^q=g_2^{-1}(Y_i)$, $i= 1,\ldots,n$, the above equations can be written as
\begin{align*}
	{\partial l(\boldsymbol{\theta},p,q) \over\partial \alpha}
	&=
	n\left[
	\log(\gamma)-\psi^{(0)}(\alpha)
	+
	{1\over n}
	\sum_{i=1}^{n}
	\log(X_i)
	\right],  
	\\[0,2cm]
	{\partial l(\boldsymbol{\theta},p,q) \over\partial \beta}
	&=
	n
	\left[
	\log(\gamma)-\psi^{(0)}(\beta)
	+
	{1\over n}
	\sum_{i=1}^{n}
	\log(Y_i-X_i)
	\right],  
	\\[0,2cm]
	{\partial l(\boldsymbol{\theta},p,q) \over\partial \gamma}
	&=
	n\left[
	{\alpha+\beta\over\gamma} 
	-
	{1\over n}
	\sum_{i=1}^{n} Y_i
	\right],
	\\[0,2cm]
	{\partial  l(\boldsymbol{\theta},p,q) \over\partial p}
	&=
	{n\over p}
	\Bigg[
	1
	+
	{1\over n}
	\sum_{i=1}^{n}
	\log(g_1^{-1}(X_i))
	+
	{1\over n}
	\sum_{i=1}^{n}
	{g_1''(g_1^{-1}(X_i)) g_1^{-1}(X_i)\log(g_1^{-1}(X_i))\over g_1'(g_1^{-1}(X_i))} 
	\\[0,2cm]
	&
	+
	(\alpha-1) \,
	{1\over n}
	\sum_{i=1}^{n}
	{g_1'(g_1^{-1}(X_i))g_1^{-1}(X_i)\log(g_1^{-1}(X_i))\over X_i} 
	\\[0,2cm]
	&
	-
	(\beta-1) \,
	{1\over n}
	\sum_{i=1}^{n}
	{g_1'(g_1^{-1}(X_i))g_1^{-1}(X_i)\log(g_1^{-1}(X_i))\over Y_i-X_i} 
	\Bigg], 
	\\[0,2cm]
	{\partial  l(\boldsymbol{\theta},p,q) \over\partial q}
	&=
	{n\over q}
	\Bigg[
	1
	+
	{1\over n}
	\sum_{i=1}^{n}
	\log(g_2^{-1}(Y_i))
	+
	{1\over n}
	\sum_{i=1}^{n}
	{g_2''(g_2^{-1}(Y_i))g_2^{-1}(Y_i)\log(g_2^{-1}(Y_i))\over g_2'(g_2^{-1}(Y_i))} 
	\\[0,2cm]
	&+
	(\beta-1) \,
	{1\over n}
	\sum_{i=1}^{n}
	{g_2'(g_2^{-1}(Y_i))g_2^{-1}(Y_i)\log(g_2^{-1}(Y_i))\over Y_i-X_i} 
	\\[0,2cm]
	&
	-
	\gamma \,
	{1\over n}
	\sum_{i=1}^{n}
	{g_2'(g_2^{-1}(Y_i))g_2^{-1}(Y_i)\log(g_2^{-1}(Y_i))}
	\Bigg].
\end{align*}
Considering the notation
\begin{align}\label{def-g}
	\begin{array}{llllll}
	&\displaystyle
	\overline{Z}_{1}\equiv{1\over n}\sum_{i=1}^{n}h_1(X_i,Y_i),
	\quad h_1(x,y)=y,
	\\[0,5cm]
	&\displaystyle
	\overline{Z}_{2}\equiv{1\over n} \sum_{i=1}^{n}h_2(X_i,Y_i),
	\quad h_2(x,y)=\log(g_1^{-1}(x)),
		\\[0,5cm]
	&\displaystyle
	\overline{Z}_{3}\equiv{1\over n} \sum_{i=1}^{n}h_3(X_i,Y_i), 
	\quad h_3(x,y)=	{g_1''(g_1^{-1}(x)) g_1^{-1}(x)\log(g_1^{-1}(x))\over g_1'(g_1^{-1}(x))},
		\\[0,5cm]
	&\displaystyle
	\overline{Z}_{4}\equiv{1\over n} \sum_{i=1}^{n}h_4(X_i,Y_i),
	\quad h_4(x,y)=	{g_1'(g_1^{-1}(x))g_1^{-1}(x)\log(g_1^{-1}(x))\over x} ,
		\\[0,5cm]
	&\displaystyle
	\overline{Z}_{5}\equiv{1\over n} \sum_{i=1}^{n}h_5(X_i,Y_i),
	\quad h_5(x,y)=	{g_1'(g_1^{-1}(x))g_1^{-1}(x)\log(g_1^{-1}(x))\over y-x},
		\\[0,5cm]
	&\displaystyle
	\overline{Z}_{6}\equiv{1\over n} \sum_{i=1}^{n}h_6(X_i,Y_i),
	\quad h_6(x,y)=\log(g_2^{-1}(y)),
		\\[0,5cm]
&\displaystyle
\overline{Z}_{7}\equiv{1\over n} \sum_{i=1}^{n}h_7(X_i,Y_i),
\quad h_7(x,y)=	{g_2''(g_2^{-1}(y))g_2^{-1}(y)\log(g_2^{-1}(y))\over g_2'(g_2^{-1}(y))},
		\\[0,5cm]
&\displaystyle
\overline{Z}_{8}\equiv{1\over n} \sum_{i=1}^{n}h_8(X_i,Y_i),
\quad h_8(x,y)=	{g_2'(g_2^{-1}(y))g_2^{-1}(y)\log(g_2^{-1}(y))\over y-x},
		\\[0,5cm]
&\displaystyle
\overline{Z}_{9}\equiv{1\over n} \sum_{i=1}^{n}h_9(X_i,Y_i),
\quad h_9(x,y)=	{g_2'(g_2^{-1}(y))g_2^{-1}(y)\log(g_2^{-1}(y))},
	\end{array}
\end{align}
the above equations containing the terms 
${\partial  l(\boldsymbol{\theta},p,q)/\partial\gamma}, {\partial  l(\boldsymbol{\theta},p,q)/\partial p}$ and ${\partial  l(\boldsymbol{\theta},p,q)/\partial q}$ may be expressed more simply as:
\begin{align}
	{\partial l(\boldsymbol{\theta},p,q) \over\partial \gamma}
&=
n\left[
{\alpha+\beta\over\gamma} 
-
\overline{Z}_{1}
\right],
\nonumber
\\[0,2cm]
{\partial  l(\boldsymbol{\theta},p,q) \over\partial p}
&=
{n\over p}
\left[
1
+
\overline{Z}_{2}
+
\overline{Z}_{3}
+
(\alpha-1) 
\overline{Z}_{4} 
-
(\beta-1) 
\overline{Z}_{5}
\right], 
\label{eq-1}
\\[0,2cm]
{\partial  l(\boldsymbol{\theta},p,q) \over\partial q}
&=
{n\over q}
\left[
1
+
\overline{Z}_{6}
+
\overline{Z}_{7}
+
(\beta-1) 
\overline{Z}_{8}
-
\gamma 
\overline{Z}_{9}
\right].
\label{eq-2}
\end{align}

By resolving  ${\partial l(\boldsymbol{\theta},p,q)/\partial \gamma}=0$, we can express $\gamma$ as a function of $(\alpha,\beta)^\top$:
%
\begin{align}\label{est-gamma}
{\gamma}(\alpha,\beta)
=
{\alpha+\beta\over \overline{Z}_1}.
\end{align}

From the equation ${\partial l(\boldsymbol{\theta},p,q)/\partial q}=0$ with $\gamma$ as in \eqref{est-gamma}, and after multiplying both sides by $\overline{Z}_{4}$, it follows that:
\begin{align}\label{eq-pre-1}
	\overline{Z}_1\overline{Z}_{4}
	+
	\overline{Z}_1\overline{Z}_{4}\overline{Z}_{6}
	+
	\overline{Z}_1\overline{Z}_{4}\overline{Z}_{7}
	-
	\overline{Z}_1\overline{Z}_{4}\overline{Z}_{8}
	-
\alpha\overline{Z}_{4}\overline{Z}_{9}
	+
\beta 
(
\overline{Z}_1\overline{Z}_{8} 
-
\overline{Z}_{9}
)
\overline{Z}_{4}
=0.
\end{align}

Multiplying ${\partial l(\boldsymbol{\theta},p,q)/\partial p}=0$ by $\overline{Z}_{9}$, we get:
\begin{align}\label{eq-pre-2}
\overline{Z}_{9}
+
\overline{Z}_{2}\overline{Z}_{9}
+
\overline{Z}_{3}\overline{Z}_{9}
-
\overline{Z}_{4}\overline{Z}_{9}
+
\overline{Z}_{5}\overline{Z}_{9}
+
\alpha
\overline{Z}_{4}\overline{Z}_{9}  
-
\beta
\overline{Z}_{5}\overline{Z}_{9}
=
0.
\end{align}
Summing the left- and right-hand sides of Equations \eqref{eq-pre-1} and \eqref{eq-pre-2} separately, we arrive at:
\begin{align}\label{est-beta}
	\widehat{\beta} 
	=
	\dfrac
	{
	(
	1
	+
	\overline{Z}_{2}
	+
	\overline{Z}_{3}
	-
	\overline{Z}_{4}
	+
	\overline{Z}_{5}
	)
	\overline{Z}_{9}
	+
	(
	1
	+
	\overline{Z}_{6}
	+
	\overline{Z}_{7}
	-
	\overline{Z}_{8}
	)
	\overline{Z}_1\overline{Z}_{4}
}{	
(
\overline{Z}_{9}
-
\overline{Z}_1\overline{Z}_{8} 
)
\overline{Z}_{4}
+
\overline{Z}_{5}\overline{Z}_{9}
}.
\end{align}

By substituting $\widehat{\beta}$ into ${\partial l(\boldsymbol{\theta},p,q)/\partial p}=0$ gives
\begin{align}\label{est-alpha}
	\widehat{\alpha} 
	=
	\dfrac{
	(\widehat{\beta} -1) 
	\overline{Z}_{5}
	-
	1
	-
	\overline{Z}_{2}
	-
	\overline{Z}_{3}
	+
	\overline{Z}_{4} 
}{	\overline{Z}_{4}}.
\end{align}

Plugging \eqref{est-beta} and \eqref{est-alpha} in \eqref{est-gamma} yields:
\begin{align}\label{est-gamma-1}
	\widehat{\gamma}
	=
	{\widehat{\alpha}+\widehat{\beta}\over \overline{Z}_1}.
\end{align}

\begin{remark}
	Let $(X,Y)^\top$  have  density given in \eqref{pdf-1}.
	Since $Y\sim{\rm Gamma}(\alpha+\beta,\gamma)$ (see lines below Item \eqref{rep-stoch}),  we have
	\begin{align}\label{qe-1}
		\mathbb{E}[h_{1}]
		\equiv
		\mathbb{E}\left[h_1(X,Y)\right]
		=
				\mathbb{E}\left[Y\right]
=
{\alpha+\beta\over\gamma}.
	\end{align}

	On the other hand, suppose that for some $p\in[a,b]$ and $q\in[c,d]$, the  partial derivatives $\partial f(u,v;\boldsymbol{\theta},p,q)/\partial p$ and $\partial f(u,v;\boldsymbol{\theta},p,q)/\partial q$
	exist on $\mathbb{R}^2\times [a,b]$ and $\mathbb{R}^2\times [c,d]$, respectively, and that there exist two integrable functions $H_1$ and $H_2$ on $\mathbb{R}^2$ such that
	\begin{align}\label{bound-unif}
	\left\vert {f(u,v;\boldsymbol{\theta},p,q)\over \partial p}\right\vert
	\leqslant 
	H_1(u,v),
		\quad 
	\left\vert {f(u,v;\boldsymbol{\theta},p,q)\over \partial q}\right\vert
	\leqslant 
	H_2(u,v).
	\end{align}
	Then, by applying the dominated convergence Theorem, we may interchange the order of integration and partial differentiation for the improper integral; that is,
	\begin{align}
		\mathbb{E}\left[{\partial  l(\boldsymbol{\theta},p,q) \over\partial p}\right]
		=
		n^2
		\mathbb{E}
		\left[{\partial \log(f(U,V;\boldsymbol{\theta},p,q))\over \partial p}\right]
		\nonumber
		&=
		n^2\left[\int\int {\partial f(u,v;\boldsymbol{\theta},p,q)\over \partial p} {\rm d}u{\rm d}v\right]
		\nonumber 
		\\[0,2cm]
		&=
		n^2\left[{\partial \over \partial p} 
		\int\int f(u,v;\boldsymbol{\theta},p,q) {\rm d}u{\rm d}v\right] 
		=0,\label{qe-2}
	\end{align}
	and, similarly,
	\begin{align}
	\mathbb{E}\left[{\partial  l(\boldsymbol{\theta},p,q) \over\partial q}\right]
	=
	0.
	\label{qe-3}
	\end{align}
	
	Thus, using the identities \eqref{qe-1}, \eqref{qe-2} and \eqref{qe-3} and taking expectations on both sides of \eqref{eq-1} and \eqref{eq-2}, we proceed analogously to derive the estimators $\widehat{\alpha}, \widehat{\beta}$ and $\widehat{\gamma}$, obtaining:
\begin{align}\label{est-alpha-exp}
	\alpha 
	=
	\dfrac{
		(\beta -1) 
		\mathbb{E}[h_{5}]
		-
		1
		-
		\mathbb{E}[h_{2}]
		-
		\mathbb{E}[h_{3}]
		+
		\mathbb{E}[h_{4}]
	}{\mathbb{E}[h_{4}]},
\end{align}
\begin{align}\label{est-beta-exp}
	{\beta} 
	=
	\dfrac
	{
		(
		1
		+
		\mathbb{E}[h_{2}]
		+
		\mathbb{E}[h_{3}]
		-
		\mathbb{E}[h_{4}]
		+
		\mathbb{E}[h_{5}]
		)
		\mathbb{E}[h_{9}]
		+
		(
		1
		+
		\mathbb{E}[h_{6}]
		+
		\mathbb{E}[h_{7}]
		-
		\mathbb{E}[h_{8}]
		)
		\mathbb{E}[h_{1}]\mathbb{E}[h_{4}]
	}{	
		(
		\mathbb{E}[h_{9}]
		-
		\mathbb{E}[h_{1}]\mathbb{E}[h_{8}]
		)
		\mathbb{E}[h_{4}]
		+
		\mathbb{E}[h_{5}]\mathbb{E}[h_{9}]
	}
\end{align}  
and
\begin{align}\label{est-gamma-1-exp}
	{\gamma}
	=
	{{\alpha}+{\beta}\over \mathbb{E}[h_{1}]},
\end{align}
where by abuse of notation we are writing $\mathbb{E}[h_{i}]$ to refer to $\mathbb{E}[h_{i}(X,Y)]$, $i=1,\ldots,9$.
\end{remark}

\subsection{Obtaining the estimators proposed by \cite{Zhao2022}}

Letting $g_1$ and $g_2$ be identity functions in \eqref{def-g}, and using the identities
\begin{align*}
	\begin{array}{llllll}
		&\displaystyle
		\overline{Z}_{1}={1\over n}\sum_{i=1}^{n}Y_i,
		&
		\displaystyle
		\overline{Z}_{2}
		=
		\overline{Z}_{4}
		=
		{1\over n} \sum_{i=1}^{n}\log(X_i),
		&
		\displaystyle
		\overline{Z}_{3}
		=
		\overline{Z}_{7}
		=
		0,
		&\displaystyle
		\overline{Z}_{5}
		=
		{1\over n} \sum_{i=1}^{n}{X_i\log(X_i)\over Y_i-X_i},
				\\[0,5cm]
		& 
		\displaystyle
		\overline{Z}_{6}
		=
		{1\over n} \sum_{i=1}^{n}\log(Y_i),
		& 
		\displaystyle
		\overline{Z}_{8}
		=
		{1\over n} \sum_{i=1}^{n}{Y_i\log(Y_i)\over Y_i-X_i},
		&
		\displaystyle
		\overline{Z}_{9}
		=
		{1\over n} \sum_{i=1}^{n}Y_i\log(Y_i),
	\end{array}
\end{align*}
from \eqref{est-alpha}, \eqref{est-beta} and \eqref{est-gamma-1} we get estimators for $\alpha$, $\beta$ and $\gamma$ as
\begin{align*}
	\widehat{\alpha} 
	=
	\dfrac{\displaystyle
		(\widehat{\beta} -1) 
	\left[{1\over n} \sum_{i=1}^{n}{X_i\log(X_i)\over Y_i-X_i}\right]
		-
		1 
	}{\displaystyle
	{1\over n} \sum_{i=1}^{n}\log(X_i)},
\end{align*}
{\small 
\begin{align*}
	\widehat{\beta} 
	=
	\dfrac
	{\displaystyle
		\left[
		1
		+
		{1\over n} \sum_{i=1}^{n}{X_i\log(X_i)\over Y_i-X_i}
		\right]
		\left[{1\over n} \sum_{i=1}^{n}Y_i\log(Y_i)\right]
		+
		\left[
		1
		+
		{1\over n} \sum_{i=1}^{n}\log(Y_i)
		-
		{1\over n} \sum_{i=1}^{n}{Y_i\log(Y_i)\over Y_i-X_i}
		\right]
		\left[{1\over n}\sum_{i=1}^{n}Y_i\right]
		\left[{1\over n} \sum_{i=1}^{n}\log(X_i)\right]
	}{\displaystyle
		\left\{
		{1\over n} \sum_{i=1}^{n}Y_i\log(Y_i)
		-
		\left[{1\over n}\sum_{i=1}^{n}Y_i\right]
		\left[{1\over n} \sum_{i=1}^{n}{Y_i\log(Y_i)\over Y_i-X_i}\right]
		\right\}
		\left[{1\over n} \sum_{i=1}^{n}\log(X_i)\right]
		+
		\left[{1\over n} \sum_{i=1}^{n}{X_i\log(X_i)\over Y_i-X_i}\right]
		\left[{1\over n} \sum_{i=1}^{n}Y_i\log(Y_i)\right]
	}
\end{align*}
}\noindent
and
\begin{align*}
	\widehat{\gamma}
	=
	\dfrac{\widehat{\alpha}+\widehat{\beta}
}{\displaystyle
	{1\over n}\sum_{i=1}^{n}Y_i},
\end{align*}
respectively, 
which coincide with those proposed by \cite{Zhao2022}.

\subsection{Proposing new closed-form estimators}\label{new_proposed}

Taking $g_1(x)=\exp(rx)-1$ and $g_2(y)=\exp(sx)-1$ in \eqref{def-g}, where $r,s>0$, and plugging in Equations \eqref{est-alpha}, \eqref{est-beta} and \eqref{est-gamma-1} the following identities
\begin{align*}
	\begin{array}{llllll}
		&\displaystyle
		\overline{Z}_{1}
		=
		{1\over n}\sum_{i=1}^{n} Y_i,
		&\displaystyle
		\overline{Z}_{2}
		=
		{1\over n} \sum_{i=1}^{n} 
		\log\left({\log(X_i+1)\over r}\right),
		\\[0,5cm]
		&\displaystyle
		\overline{Z}_{3}
		=
		{1\over n} 
		\sum_{i=1}^{n}
		{\log(X_i+1)\log\left({\log(X_i+1)\over r}\right)},
		&\displaystyle
		\overline{Z}_{4}
		=
		{1\over n} 
		\sum_{i=1}^{n}	{(X_i+1)\log(X_i+1)\log\left({\log(X_i+1)\over r}\right)\over X_i} ,
		\\[0,5cm]
		&\displaystyle
		\overline{Z}_{5}
		=
		{1\over n} 
		\sum_{i=1}^{n}	{(X_i+1)\log(X_i+1)\log\left({\log(X_i+1)\over r}\right)\over Y_i-X_i},
		&\displaystyle
		\overline{Z}_{6}
		=
		{1\over n} 
		\sum_{i=1}^{n}\log\left({\log(Y_i+1)\over s}\right),
		\\[0,5cm]
		&\displaystyle
		\overline{Z}_{7}
		=
		{1\over n} 
		\sum_{i=1}^{n}	{\log(Y_i+1)\log\left({\log(Y_i+1)\over s}\right)},
		&\displaystyle
		\overline{Z}_{8}
		=
		{1\over n} 
		\sum_{i=1}^{n}	{(Y_i+1)\log(Y_i+1)\log\left({\log(Y_i+1)\over s}\right)\over Y_i-X_i},
		\\[0,5cm]
		&\displaystyle
		\overline{Z}_{9}
		=
		{1\over n} 
		\sum_{i=1}^{n}	
		{ (Y_i+1)\log(Y_i+1)\log\left({\log(Y_i+1)\over s}\right)},
	\end{array}
\end{align*}
we get new closed-form estimators for $\alpha$, $\beta$ and $\gamma$, respectively.
\begin{remark}
	As the estimators above depend on $r>0$ and
 	$s>0$, different choices of these parameters yield multiple estimators for $\alpha$, $\beta$ and $\gamma$.
 \end{remark}

In the next section, we show that the new estimator $(\widehat{\alpha},\widehat{\beta},\widehat{\gamma})^\top$ of $({\alpha},{\beta},{\gamma})^\top$, established in \eqref{est-alpha}, \eqref{est-beta} and \eqref{est-gamma-1},  is strongly consistent and normally asymptotic.

\section{Large sample properties}\label{largeprop}

Let $\{(X_i,Y_i)^{\top} : i = 1,\ldots , n\}$ be a bivariate random sample (independent and identically distributed) of size $n$ from  $(X,Y)^{\top}$ with PDF \eqref{pdf-1}. 

As $h_i, i=1,\ldots,9,$ in \eqref{def-g} are measurable functions, the sequence of 9-dimensional random vectors  
\begin{align}\label{vectors-h}
	\begin{pmatrix}
		h_1(X_1,Y_1)\\
		h_2(X_1,Y_1)\\
		h_3(X_1,Y_1)\\
		h_4(X_1,Y_1)\\
		h_5(X_1,Y_1)\\
		h_6(X_1,Y_1)\\
				h_7(X_1,Y_1)\\  
						h_8(X_1,Y_1)\\  
								h_9(X_1,Y_1)    
	\end{pmatrix},
	\ldots,
	\begin{pmatrix}
		h_1(X_n,Y_n)\\
		h_2(X_n,Y_n)\\
		h_3(X_n,Y_n)\\
		h_4(X_n,Y_n)\\
		h_5(X_n,Y_n)\\
		h_6(X_n,Y_n)\\ 
				h_7(X_n,Y_n)\\  
						h_8(X_n,Y_n)\\  
								h_9(X_n,Y_n) 
	\end{pmatrix}
\end{align}
are independent and identically distributed with 
\begin{align}\label{def-vector-Z}
	\boldsymbol{Z}
	\equiv
	\begin{pmatrix}
	h_1(X,Y)\\
	h_2(X,Y)\\
	h_3(X,Y)\\
	h_4(X,Y)\\
	h_5(X,Y)\\
	h_6(X,Y)\\  
	h_7(X,Y)\\  
	h_8(X,Y)\\  
	h_9(X,Y) 
\end{pmatrix}.
\end{align}
So, by applying strong law of large
numbers, we have
\begin{align*}
	\overline{\boldsymbol{Z}}
	\equiv
	\begin{pmatrix}
		\overline{Z}_1\\
		\overline{Z}_2\\
		\overline{Z}_3\\
		\overline{Z}_4\\
		\overline{Z}_5\\
		\overline{Z}_6\\ 
				\overline{Z}_7\\  
						\overline{Z}_8\\  
								\overline{Z}_9\\   
	\end{pmatrix}
	\stackrel{\rm a.s.}{\longrightarrow}
	\mathbb{E}[\boldsymbol{Z}]
	=
		\begin{pmatrix}
		\mathbb{E}[h_1(X,Y)]\\
		\mathbb{E}[h_2(X,Y)]\\
		\mathbb{E}[h_3(X,Y)]\\
		\mathbb{E}[h_4(X,Y)]\\
		\mathbb{E}[h_5(X,Y)]\\
		\mathbb{E}[h_6(X,Y)]\\ 
				\mathbb{E}[h_7(X,Y)]\\  
						\mathbb{E}[h_8(X,Y)]\\  
								\mathbb{E}[h_9(X,Y)]\\   
	\end{pmatrix}
	\equiv 
			\begin{pmatrix}
		\mathbb{E}[h_1]\\
		\mathbb{E}[h_2]\\
		\mathbb{E}[h_3]\\
		\mathbb{E}[h_4]\\
		\mathbb{E}[h_5]\\
		\mathbb{E}[h_6]\\ 
		\mathbb{E}[h_7]\\  
		\mathbb{E}[h_8]\\  
		\mathbb{E}[h_9]\\   
	\end{pmatrix},
\end{align*}
where $\overline{Z}_i$, $i=1,\ldots,9$, are given in \eqref{def-g} and ``$\stackrel{\rm a.s.}{\longrightarrow}$'' denotes almost sure convergence.

Using notation 
\begin{align}\label{def-h1}
	&\xi_1(\boldsymbol{z})
	\equiv
	\dfrac{
		[\xi_2(\boldsymbol{z}) -1]
		z_{5}
		-
		1
		-
		z_{2}
		-
		z_{3}
		+
		z_{4} 
	}{z_{4}},
	\\[0,2cm]
	&
	\xi_2(\boldsymbol{z})
	\equiv
	\dfrac
	{
		(
		1
		+
		z_{2}
		+
		z_{3}
		-
		z_{4}
		+
		z_{5}
		)
		z_{9}
		+
		(
		1
		+
		z_{6}
		+
		z_{7}
		-
		z_{8}
		)
		z_1z_{4}
	}{	
		(
		z_{9}
		-
		z_1z_{8} 
		)
		z_{4}
		+
		z_{5}z_{9}
	}
\end{align}
and
\begin{align}\label{def-h2}
	\xi_3(\boldsymbol{z})
	\equiv 
	{\xi_1(\boldsymbol{z})+\xi_2(\boldsymbol{z})\over z_1},
\end{align}
from \eqref{est-alpha}, \eqref{est-beta} and \eqref{est-gamma-1} we have
\begin{align}\label{id-0}
	\begin{pmatrix}
		\widehat{\alpha}\\
		\widehat{\beta}\\
		\widehat{\gamma}
	\end{pmatrix}
	=
	\begin{pmatrix}
		\xi_1(\overline{\boldsymbol{Z}})\\
		\xi_2(\overline{\boldsymbol{Z}})\\
		\xi_3(\overline{\boldsymbol{Z}})
	\end{pmatrix}.
\end{align}
Hence, continuous-mapping theorem gives
\begin{align}\label{id-1}
		\begin{pmatrix}
	\widehat{\alpha}\\
	\widehat{\beta}\\
		\widehat{\gamma}
		\end{pmatrix}
	\stackrel{\eqref{id-0}}{=}
		\begin{pmatrix}
	\xi_1(\overline{\boldsymbol{Z}})\\
	\xi_2(\overline{\boldsymbol{Z}})\\
	\xi_3(\overline{\boldsymbol{Z}})
		\end{pmatrix}
	\stackrel{\rm a.s.}{\longrightarrow}
			\begin{pmatrix}
	\xi_1(\mathbb{E}[\boldsymbol{Z}])\\
	\xi_2(\mathbb{E}[\boldsymbol{Z}])\\
	\xi_3(\mathbb{E}[\boldsymbol{Z}])\\
			\end{pmatrix}.
\end{align}

Moreover, since the random vectors in \eqref{vectors-h}  are independent and identically distributed with $\boldsymbol{Z}$, by multivariate central limit theorem (CLT) we obtain
\begin{align*}
	\sqrt{n}\big\{\overline{\boldsymbol{Z}}-\mathbb{E}[\boldsymbol{Z}]\big\}\stackrel{\mathscr D}{\longrightarrow} 
	N_9\left(
		\begin{pmatrix}
		0\\
		0\\
		0\\
		0\\
		0\\
		0\\
				0\\
						0\\
								0
	\end{pmatrix}, 
	\bm\Sigma_{_{\boldsymbol{Z}}}\right),
\end{align*}
	where ``$\stackrel{\mathscr D}{\longrightarrow}$'' means convergence in distribution and $\bm\Sigma_{_{\boldsymbol{Z}}}$ denotes the covariance matrix of $\boldsymbol{Z}$.
So, delta method provides
\begin{align}\label{id-2}
	\sqrt{n}
	\left\{
	\begin{pmatrix}
		\widehat{\alpha}
		\\
		\widehat{\beta}
		\\
		\widehat{\gamma}	
	\end{pmatrix}
	-
	\begin{pmatrix}
		\xi_1(\mathbb{E}[\bm Z])
		\\
		\xi_2(\mathbb{E}[\bm Z])
				\\
		\xi_3(\mathbb{E}[\bm Z])
	\end{pmatrix}
	\right\}
		\stackrel{\eqref{id-0}}{=}
		\sqrt{n}
		\left\{
		\begin{pmatrix}
			\xi_1(\overline{\boldsymbol{Z}})
			\\
			\xi_2(\overline{\boldsymbol{Z}})
						\\
			\xi_3(\overline{\boldsymbol{Z}})
		\end{pmatrix}
		-
		\begin{pmatrix}
			\xi_1(\mathbb{E}[\bm Z])
			\\
			\xi_2(\mathbb{E}[\bm Z])
					\\
			\xi_3(\mathbb{E}[\bm Z])
		\end{pmatrix}
		\right\}
	\stackrel{\mathscr D}{\longrightarrow}
	N_3\left(		
	\begin{pmatrix}
		0\\
		0\\
		0
	\end{pmatrix}, 
	\bm A\bm\Sigma_{_{\boldsymbol{Z}}} \bm A^\top
	\right),
\end{align}
with $\bm A$ being the partial derivatives matrix defined as
\begin{align}\label{matrix-A}
	\bm A
	=
	\left. 
	\begin{pmatrix}
		\displaystyle
		{\partial \xi_1(\bm z)\over\partial z_1} & \displaystyle {\partial \xi_1(\bm z)\over\partial z_2} & \displaystyle {\partial \xi_1(\bm z)\over\partial z_3} & \displaystyle {\partial \xi_1(\bm z)\over\partial z_4} & \displaystyle {\partial \xi_1(\bm z)\over\partial z_5} & \displaystyle
		{\partial \xi_1(\bm z)\over\partial z_6} & \displaystyle
		{\partial \xi_1(\bm z)\over\partial z_7} & \displaystyle
		{\partial \xi_1(\bm z)\over\partial z_8} & \displaystyle
		{\partial \xi_1(\bm z)\over\partial z_9}
		\\[0,5cm]
		\displaystyle
		{\partial \xi_2(\bm z)\over\partial z_1} & \displaystyle {\partial \xi_2(\bm z)\over\partial z_2} & \displaystyle {\partial \xi_2(\bm z)\over\partial z_3} & \displaystyle {\partial \xi_2(\bm z)\over\partial z_4} & \displaystyle {\partial \xi_2(\bm z)\over\partial z_5} & \displaystyle
		{\partial \xi_2(\bm z)\over\partial z_6} & \displaystyle
		{\partial \xi_2(\bm z)\over\partial z_7} & \displaystyle
		{\partial \xi_2(\bm z)\over\partial z_8} & \displaystyle
		{\partial \xi_2(\bm z)\over\partial z_9} 
			\\[0,5cm]
	\displaystyle
	  {\partial \xi_3(\bm z)\over\partial z_1} & \displaystyle {\partial \xi_3(\bm z)\over\partial z_2} & \displaystyle {\partial \xi_3(\bm z)\over\partial z_3} & \displaystyle {\partial \xi_3(\bm z)\over\partial z_4} & \displaystyle {\partial \xi_3(\bm z)\over\partial z_5} & \displaystyle {\partial \xi_3(\bm z)\over\partial z_6} & \displaystyle
	  {\partial \xi_3(\bm z)\over\partial z_7} & \displaystyle
	  {\partial \xi_3(\bm z)\over\partial z_8} & \displaystyle
	  {\partial \xi_3(\bm z)\over\partial z_9} 
	\end{pmatrix}\,
	\right\vert_{\bm z=\mathbb{E}(\bm Z)}.
\end{align}
For simplicity of presentation, we do not present the partial derivatives of $\xi_j$, $j=1,2,3$, nor the elements of the covariance matrix $\bm\Sigma_{_{\boldsymbol{Z}}}$, as the existence of this last matrix depends on the choice of the monotonic transformations $g_1$ and $g_2$ in \eqref{dist-gen-exp}.

Moreover, by \eqref{est-alpha-exp}, \eqref{est-beta-exp} and \eqref{est-gamma-1-exp} it follows that
\begin{align}\label{eq-exp}
		\begin{pmatrix}
		\xi_1(\mathbb{E}[\bm Z])
		\\
		\xi_2(\mathbb{E}[\bm Z])
		\\
		\xi_3(\mathbb{E}[\bm Z])
	\end{pmatrix}
	=
		\begin{pmatrix}
		\alpha
		\\
		\beta
		\\
		\gamma
	\end{pmatrix}.
\end{align}

By combining \eqref{id-1}, \eqref{id-2}, and \eqref{eq-exp}, we obtain the following theorem.
\begin{theorem}[Strong consistency and asymptotic normality]
	Under the assumptions given in \eqref{bound-unif}, we have:
\begin{align*}
	\begin{pmatrix}
		\widehat{\alpha}\\
		\widehat{\beta}\\
		\widehat{\gamma}
	\end{pmatrix}
	\stackrel{\rm a.s.}{\longrightarrow}
	\begin{pmatrix}
		\alpha\\
		\beta\\
		\gamma\\
	\end{pmatrix}
\end{align*}
and
\begin{align*}
	\sqrt{n}
	\left\{
	\begin{pmatrix}
		\widehat{\alpha}
		\\
		\widehat{\beta}
		\\
		\widehat{\gamma}	
	\end{pmatrix}
	-
	\begin{pmatrix}
		\alpha
		\\
		\beta
		\\
		\gamma
	\end{pmatrix}
	\right\}
	\stackrel{\mathscr D}{\longrightarrow}
	N_3\left(		
	\begin{pmatrix}
		0\\
		0\\
		0
	\end{pmatrix}, 
	\bm A\bm\Sigma_{_{\boldsymbol{Z}}} \bm A^\top
	\right),
\end{align*}
whenever the covariance matrix $\bm \Sigma_{_{\boldsymbol{Z}}}$ of the random vector $\boldsymbol{Z}$, defined in \eqref{def-vector-Z}, exists. Here, $\bm A$ is the $3\times 9$ matrix given in \eqref{matrix-A}.
\end{theorem}

\section{New closed-form estimators: Second approach}\label{The New Estimators-1}

It is clear that  $X$ and $Y$ in \eqref{rep-stoch} satisfies ${X/Y}\sim{\rm Beta}(\alpha,\beta)$. 
%
Then the closed-form estimators of these parameters are \citep[see Subsection 4.1 of][]{Vila2025}
\begin{align}\label{defab-1}
	\widehat{\alpha}
	=
	\dfrac
	{
		{1+\overline{U}_{3,1}+\overline{U}_{4,1}+\overline{U}_{5,1}\over \overline{U}_{2,1}}\, \overline{U}_{2,2}
		-
		1-\overline{U}_{3,2}-\overline{U}_{4,2}-\overline{U}_{5,2}
	}
	{
		\overline{U}_{1,2}-{\overline{U}_{1,1}\overline{U}_{2,2}\over \overline{U}_{2,1}}
	}
\end{align}
and
\begin{align}\label{defab-0}
	\widehat{\beta}
	=
	\dfrac{\widehat{\alpha}\overline{U}_{1,1}+1+\overline{U}_{3,1}+\overline{U}_{4,1}+\overline{U}_{5,1}}{\overline{U}_{2,1}},
\end{align}
respectively. In the above, for two univariate and independent random samples $\{X_i:i=1,\ldots,n\}$ and $\{Y_i:i=1,\ldots,n\}$ from $X$ and $Y$, respectively, the notation below is used:
\begin{align*}
	\overline{U}_{k,j}
	\equiv
	{1\over n}\sum_{i=1}^{n} 
	\vartheta_{k,j}\left({X_i\over Y_i}\right), 
	\quad k=1,2,3,4,5, \, j=1,2,
\end{align*}
 and
\begin{align}\label{defab2}
	\begin{array}{lll}
		&\displaystyle
		\vartheta_{1,j}(x)\equiv	{1\over x} \, 
		\ell_j'(\ell_j^{-1}(x)) \ell_j^{-1}(x)\log(\ell_j^{-1}(x)),
		\\[0,5cm]
		&\displaystyle
		\vartheta_{2,j}(x)\equiv	{1\over 1-x} \, 
		\ell_j'(\ell_j^{-1}(x)) \ell_j^{-1}(x)\log(\ell_j^{-1}(x)),
		\\[0,5cm]
		&\displaystyle
		\vartheta_{3,j}(x)\equiv	\frac{\ell_j''(\ell_j^{-1}(x))}{\ell_j'(\ell_j^{-1}(x))}\, \ell_j^{-1}(x) \log(\ell_j^{-1}(x)),
		\\[0,5cm]
		&\displaystyle
		\vartheta_{4,j}(x)\equiv	{(2x-1) \ell_j'(\ell_j^{-1}(x))\over x(1-x)} \, \ell_j^{-1}(x) \log(\ell_j^{-1}(x)),
		\\[0,5cm]
		&\displaystyle
		\vartheta_{5,j}(x)\equiv\log(\ell_j^{-1}(x)),
	\end{array} 	 
\end{align}
where the functions $\ell_j:D\subset (0,\infty)\to \text{supp}(X/Y)=(0,1)$, $j=1,2$, are monotonic and at least twice differentiable \citep{Vila2025}.

Combining results from \eqref{defab-1}, \eqref{defab-0} and \eqref{est-gamma}, we arrive at the following closed-form estimator for the parameter $\gamma$:
\begin{align}\label{eqqgamma}
	\widehat{\gamma}
	=
	{1\over \overline{Z}_1}
	\left[
	\dfrac
{
	{1+\overline{U}_{3,1}+\overline{U}_{4,1}+\overline{U}_{5,1}\over \overline{U}_{2,1}}\, \overline{U}_{2,2}
	-
	1-\overline{U}_{3,2}-\overline{U}_{4,2}-\overline{U}_{5,2}
}
{
	\overline{U}_{1,2}-{\overline{U}_{1,1}\overline{U}_{2,2}\over \overline{U}_{2,1}}
}
	\left(
	1
	+
	{\overline{U}_{1,1}\over \overline{U}_{2,1}}
	\right)
	+
	\dfrac{1+\overline{U}_{3,1}+\overline{U}_{4,1}+\overline{U}_{5,1}}{\overline{U}_{2,1}}
	\right],
\end{align}
where $\overline{Z}_1$ is as defined in \eqref{def-g}.

\subsection{Obtaining the estimators proposed by \cite{Nawa2023}}\label{sec:Nawa2023}

Choosing $\ell_1:D=(0,\infty)\to \text{supp}(X/Y)=(0,1)$ as $\ell_1(x)=x/(x+1)$, and $\ell_2:D=(0,1)\to \text{supp}(X/Y)=(0,1)$ where  $\ell_2(x)=-\log(x)/[1-\log(x)]$,  and using the definitions in \eqref{defab2},  
along with the identities
\begin{align*}
	\ell_1^{-1}(x)={x\over 1-x},
	\quad
	\ell_1'(\ell_1^{-1}(x))=(x-1)^2,
	\quad 
	\frac{\ell_1''(\ell_1^{-1}(x))}{\ell_1'(\ell_1^{-1}(x))}=2(x-1),
\end{align*}	
and 
\begin{align*}
	\ell_2^{-1}(x)=\exp\left({x\over x-1}\right),
	\quad
	\ell_2'(\ell_2^{-1}(x))=-(x-1)^2\exp\left({x\over 1-x}\right),
	\quad
	\frac{\ell_2''(\ell_2^{-1}(x))}{\ell_2'(\ell_2^{-1}(x))}=-(2x-1)\exp\left({x\over 1-x}\right),
\end{align*}
from Equations \eqref{defab-1}, \eqref{defab-0} and \eqref{eqqgamma}, the estimators for $\alpha$, $\beta$ and $\gamma$ are obtained as:
\begin{align*}
	\widehat{\alpha}
	=
	\dfrac{\displaystyle	
		{1\over n}\sum_{i=1}^{n}{X_i\over Y_i}}{
		\displaystyle	
		{1\over n}\sum_{i=1}^{n} {X_i\over Y_i}\log\left({X_i\over Y_i-X_i}\right)
		-
		\left[{1\over n}\sum_{i=1}^{n}{X_i\over Y_i}\right]
		\left[{1\over n}\sum_{i=1}^{n} \log\left({X_i\over Y_i-X_i}\right) \right]
	},
\end{align*}
\begin{align*}
	\widehat{\beta}
	=
	\dfrac{\displaystyle	
		1-{1\over n}\sum_{i=1}^{n}{X_i\over Y_i}}{
		\displaystyle	
		{1\over n}\sum_{i=1}^{n} {X_i\over Y_i}\log\left({X_i\over Y_i-X_i}\right)
		-
		\left[{1\over n}\sum_{i=1}^{n}{X_i\over Y_i}\right]
		\left[{1\over n}\sum_{i=1}^{n} \log\left({X_i\over Y_i-X_i}\right) \right]
	}
\end{align*} 
and
\begin{align*}
\widehat{\gamma}
=
	\dfrac{1}{\displaystyle	
\left[{1\over n}\sum_{i=1}^{n} Y_i\right]
\left\{
	{1\over n}\sum_{i=1}^{n} {X_i\over Y_i}\log\left({X_i\over Y_i-X_i}\right)
	-
	\left[{1\over n}\sum_{i=1}^{n}{X_i\over Y_i}\right]
	\left[{1\over n}\sum_{i=1}^{n} \log\left({X_i\over Y_i-X_i}\right) \right]
	\right\}
},
\end{align*}
respectively, which are identical to those proposed by \cite{Nawa2023}.

\subsection{Proposing new closed-form estimators}\label{new_proposed-1}

	Letting  $\ell_1:D=(0,1)\to \text{supp}(X/Y)=(0,1)$ be the identity function, that is, $\ell_1(x)=x$, and $\ell_2:D=(0,1)\to \text{supp}(X/Y)=(0,1)$ be defined by $\ell_2(x)=(1-x^{1/s})^{1/r}$ where $r,s>0$, and using the definitions in \eqref{defab2}, along with the identities
\begin{align*}
	\ell_1^{-1}(x)=x,
	\quad
	\ell_1'(\ell_1^{-1}(x))=1,
	\quad 
	\frac{\ell_1''(\ell_1^{-1}(x))}{\ell_1'(\ell_1^{-1}(x))}=0,
\end{align*}
and
\begin{align*}
	\ell_2^{-1}(x)=(1-x^r)^s,
	\quad
	\ell_2'(\ell_2^{-1}(x))=-{(1-x^r)^{1-s} x^{1-r}\over rs},
	\quad
	\frac{\ell_2''(\ell_2^{-1}(x))}{\ell_2'(\ell_2^{-1}(x))}=
	{(1-x^r)^{-s} x^{-r}[r-1+(1-rs)x^r]\over rs},
\end{align*}
by applying Equations \eqref{defab-1}, \eqref{defab-0}, and \eqref{eqqgamma}, we obtain the following estimators for 
$\alpha$, $\beta$ and $\gamma$:
\begin{align}\label{prop1:eq1}
	\widehat{\alpha}
	=
	{(\widehat{\beta}-1) B-1 \over A},
\end{align}
\begin{align}\label{prop1:eq2}
	\widehat{\beta}
	=
	{-r-E-C+D-rsF-{(B+1)C\over A}\over D-{BC\over A}}
\end{align}
and
\begin{align}\label{prop1:eq3}
	\widehat{\gamma}
	={\widehat{\alpha}+\widehat{\beta}\over \displaystyle	
		{1\over n}\sum_{i=1}^{n} Y_i},
\end{align}
respectively, 
where the following definitions have been adopted:
\begin{align*}
	A&\equiv {1\over n}
	\sum_{i=1}^{n}
	\log\left({X_i\over Y_i}\right),
	\\[0,2cm]
	B&\equiv{1\over n}
	\sum_{i=1}^{n}
	{X_i \over Y_i-X_i} \, 
	\log\left({X_i\over Y_i}\right),
	\\[0,2cm]
	C&\equiv{1\over n}
	\sum_{i=1}^{n}
	{Y_i^r-X_i^r\over X_i^r}  \, 
	\log\left({Y_i^r-X_i^r\over Y_i^r}\right),
	\\[0,2cm]
	D&\equiv{1\over n}
	\sum_{i=1}^{n}
	{Y_i^r-X_i^r\over (Y_i-X_i) X_i^{r-1}} \,
	\log\left({Y_i^r-X_i^r\over Y_i^r}\right),
	\\[0,2cm]
	E&\equiv 
	{1\over n}
	\sum_{i=1}^{n}
	{\left(r-{Y_i^r-X_i^r\over Y_i^r}-rs \, {X_i^r\over Y_i^r}\right) \left({Y_i^r-X_i^r\over Y_i^r}\right)^s\over \left({X_i\over Y_i}\right)^{2s-r}} \,
	\log\left({Y_i^r-X_i^r\over Y_i^r}\right),
	\\[0,2cm]
	F&\equiv	{1\over n}
	\sum_{i=1}^{n}
	\log\left({Y_i^r-X_i^r\over Y_i^r}\right).
\end{align*}

\begin{remark}
	As in Section \ref{largeprop}, the same reasoning leads to the strong consistency and asymptotic normality of the estimators for $\alpha$, $\beta$
	and $\gamma$ of this section.
\end{remark}


\section{Simulation study}\label{sec:simulation}

In this section, we examine the finite-sample performance of the proposed closed-form estimators (Sections \ref{new_proposed} and \ref{new_proposed-1}) for $(\alpha,\beta,\gamma)^\top$. For comparison, the maximum likelihood (ML) estimators, the closed-form estimators of \cite{Zhao2022},
and the closed-form estimators of \cite{Nawa2023}, are also included.

Note that the estimators introduced in Sections \ref{new_proposed} (Proposed 1) and \ref{new_proposed-1} (Proposed 2) are indexed by auxiliary parameters $r>0$ and $s>0$.
Consequently, a family of closed-form estimators $(\widehat\alpha_{r,s},\widehat\beta_{r,s},\widehat\gamma_{r,s})$ can be obtained from \eqref{est-alpha-exp}, \eqref{est-beta-exp} and \eqref{est-gamma-1-exp}, and from \eqref{prop1:eq1}, \eqref{prop1:eq2} and \eqref{prop1:eq3}, for different choices of $(r,s)$.
In this context, we propose a data-driven selection of $(r,s)$ through a profile likelihood approach.
Specifically, for each grid point $(r,s)\in\mathcal R\times\mathcal S$, we compute the corresponding closed-form estimates
$(\widehat\alpha_{r,s},\widehat\beta_{r,s},\widehat\gamma_{r,s})$ and evaluate the bivariate gamma log-likelihood $\ell(\widehat\alpha_{r,s},\widehat\beta_{r,s},\widehat\gamma_{r,s}\mid \,\bm x \,, \bm y)$.
The selected pair
$$
(\widehat r,\widehat s)\;\in\;\arg\max_{r\in\mathcal R,\;s\in\mathcal S}\;
\ell\!\left(\widehat\alpha_{r,s},\widehat\beta_{r,s},\widehat\gamma_{r,s}\,\middle|\,\bm x \,, \bm y\right)
$$
defines the proposed estimators
\begin{equation}\label{proposed_estimator}
(\widehat\alpha,\widehat\beta,\widehat\gamma)=(\widehat\alpha_{\widehat r,\widehat s},\widehat\beta_{\widehat r,\widehat s},\widehat\gamma_{\widehat r,\widehat s}).
\end{equation}
For instance, one may take
$\mathcal R=\mathcal S=\{0.1,0.2,\ldots,2.5\}$.
By profiling the likelihood over $(r,s)$, the procedure automatically selects the member of the class that maximizes the implied log-likelihood on the grid.

\begin{figure}[!ht]
    \centering
    \includegraphics[width=0.45\textwidth]{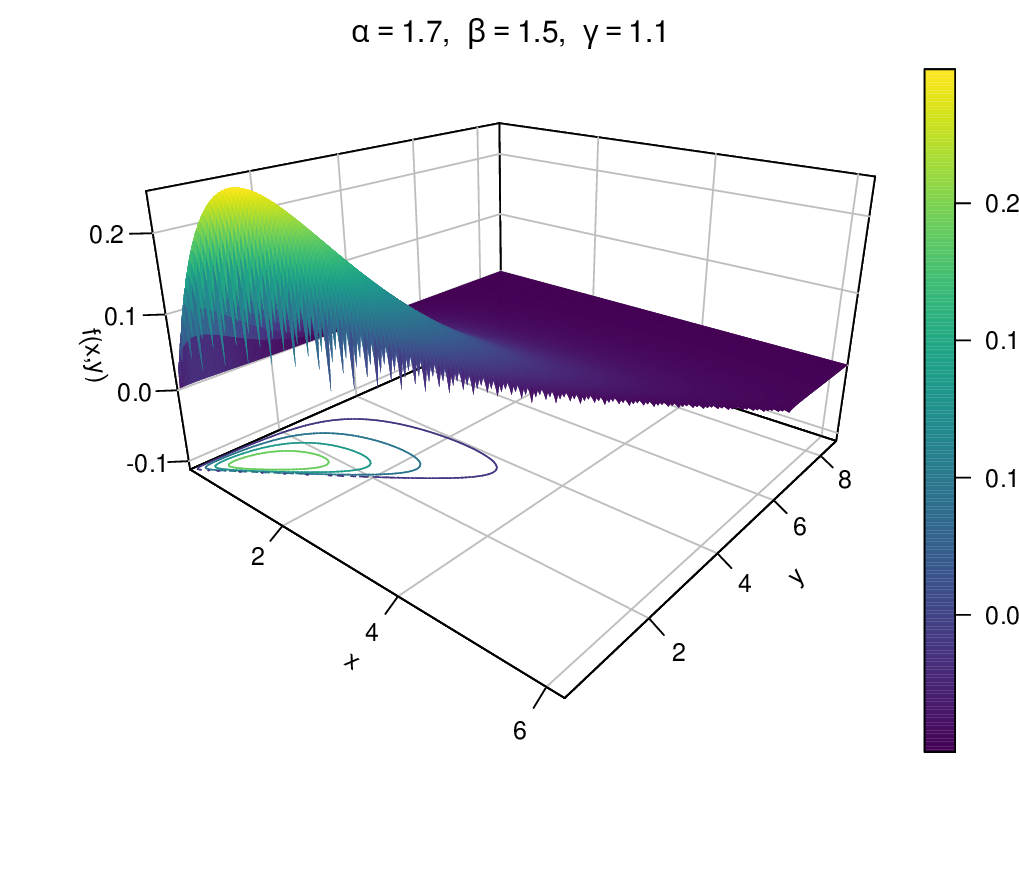}
    \includegraphics[width=0.45\textwidth]{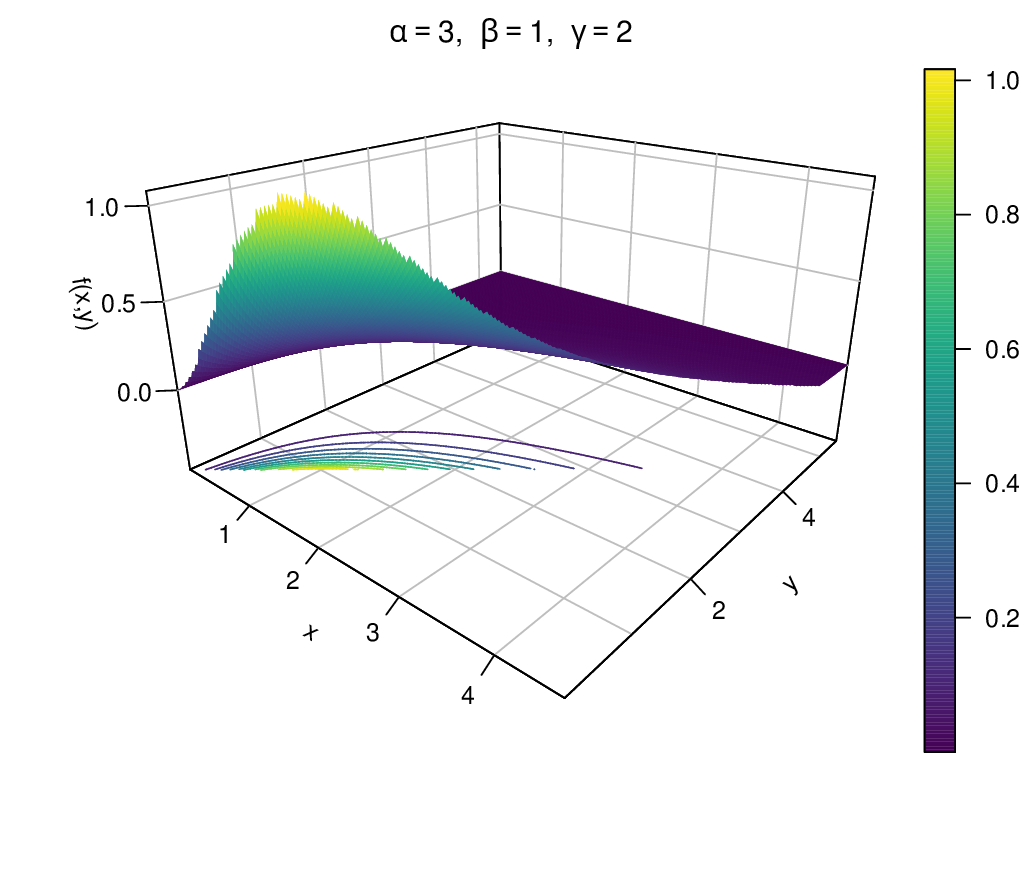}
    \caption{Shapes of the McKay’s bivariate gamma density for parameter settings
    (left) $(\alpha,\beta,\gamma)=(1.7,1.5,1.1)$ and
    (right) $(\alpha,\beta,\gamma)=(3.0,1.0,2.0)$.}
    \label{fig:shape1}
\end{figure}

\begin{figure}[!ht]
    \centering
    \includegraphics[width=0.45\textwidth]{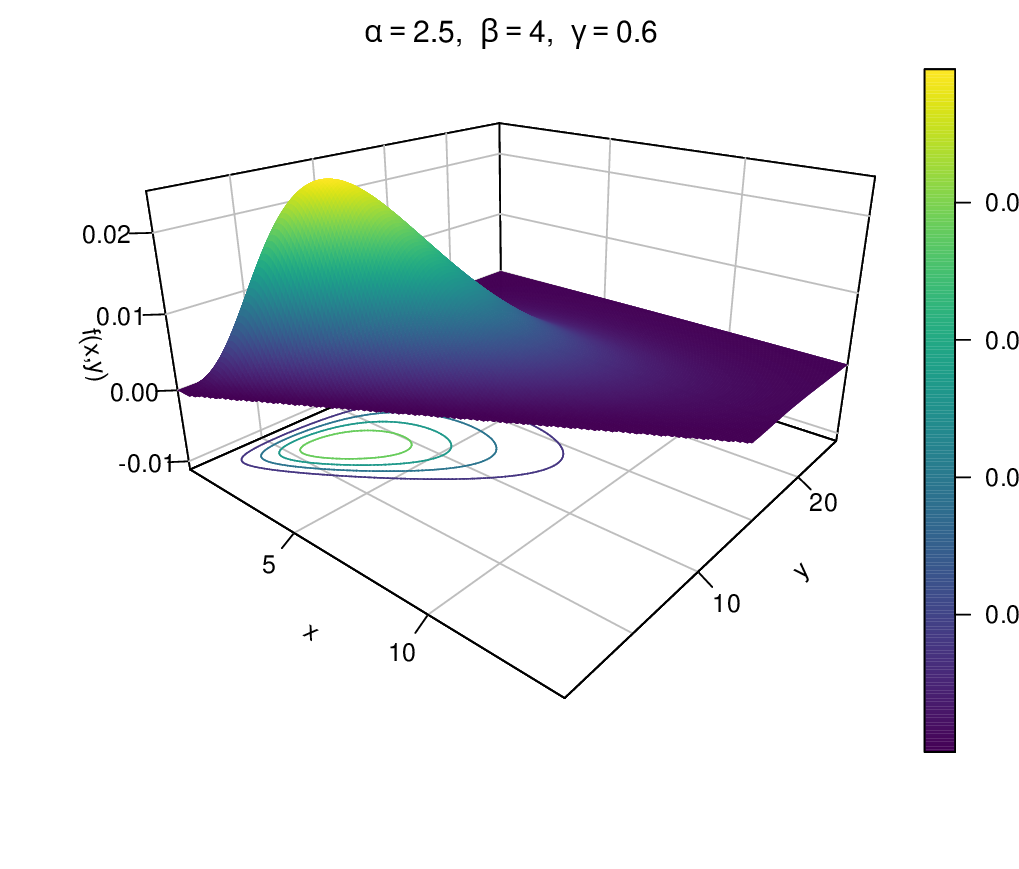}
    \includegraphics[width=0.45\textwidth]{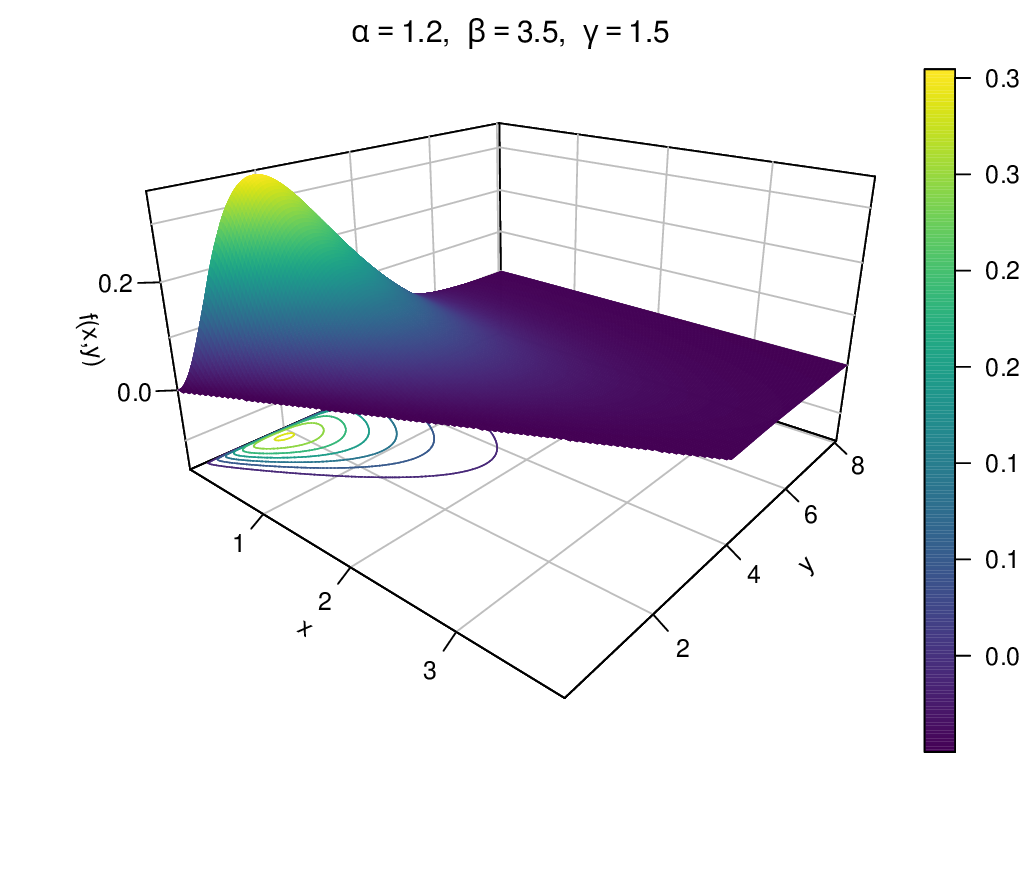}
    \caption{Shapes of the McKay’s bivariate gamma density for parameter settings
    (left) $(\alpha,\beta,\gamma)=(2.5,4.0,0.6)$ and
    (right) $(\alpha,\beta,\gamma)=(1.2,3.5,1.5)$.}
    \label{fig:shape2}
\end{figure}

We considered four scenarios for the true parameters:
$$
(\alpha,\beta,\gamma) \in \{(1.7,1.5,1.1), (3.0,1.0,2.0), (2.5,4.0,0.6), (1.2,3.5,1.5)\}.
$$
Figures~\ref{fig:shape1}--\ref{fig:shape2} display the three-dimensional surfaces of the joint density for each of the four scenarios used in the simulation study.
These plots highlight how different combinations of $(\alpha,\beta,\gamma)$ influence the skewness, tail behavior, and dependence structure between $X$ and $Y$.

For each configuration, independent random samples of sizes $n \in \{20,50,100\}$ were generated using the stochastic representation $(X,Y)^\top = (X_1, X_1+X_2)^\top$, where $X_1 \sim \text{Gamma}(\alpha,\gamma)$ and $X_2 \sim \text{Gamma}(\beta,\gamma)$ are independent. Each Monte Carlo experiment was replicated $M=1000$ times. To assess estimator performance, we considered three measures: absolute bias (AB), mean absolute relative error (MARE), and root mean square error (RMSE). These are defined, for a generic parameter $\theta$ with estimates $\hat\theta_1,\dots,\hat\theta_M$, as
\begin{eqnarray*}
\text{AB} &=& \left| \frac{1}{M}\sum_{j=1}^M \hat\theta_j - \theta \right|, \\
\text{MARE} &=& \frac{1}{M}\sum_{j=1}^M \frac{|\hat\theta_j - \theta|}{\theta}, \\
\text{RMSE} &=& \sqrt{\frac{1}{M}\sum_{j=1}^M (\hat\theta_j - \theta)^2}.
\end{eqnarray*}

The simulation results are summarized in Tables~\ref{tab:sim1}--\ref{tab:sim4}. Several important conclusions can be drawn:
\begin{enumerate}
\item The proposed closed-form estimators, Proposed~1 (Sec.~\ref{new_proposed}, profile over $(r,s)$) and Proposed~2 (Sec.~\ref{new_proposed-1}, profile over $(r,s)$), exhibit small bias and RMSE across all scenarios, closely matching the performance of ML even for small $n$.
\item Across settings, the two proposed estimators present similar performance in terms of RMSE; when differences appear, they are minor and diminish as $n$ increases.
\item The estimators of \cite{Zhao2022} generally present larger bias and RMSE.
\item The estimators of \cite{Nawa2023} are competitive but tends to display larger bias and RMSE than the proposed methods in our simulations.
\item As expected, accuracy improves with increasing $n$, supporting the consistency results in Section~\ref{largeprop}.
\item The profile-based choice of $(r,s)$ yields numerically stable estimates with no convergence issues in our runs, while avoiding the iterative optimization required by ML.
\end{enumerate}

These findings highlight that the proposed closed-form estimators achieve nearly the same efficiency as ML while avoiding iterative optimization. Furthermore, they consistently outperform the existing closed-form approaches of \cite{Zhao2022} and \cite{Nawa2023}, making them an attractive alternative in practice.

\begin{table}[!ht]
\centering
\vskip -1.5cm
\caption{Monte Carlo summary for scenario $(\alpha,\beta,\gamma)=(1.7,1.5,1.1)$. Reported metrics per sample size $n$ and estimator.}\label{tab:sim1}
\begin{tabular}{rllrrr}
\toprule
$n$ & Method  & Estimator& AB & MARE & RMSE\\
\midrule
20 & Proposed 1 & $\widehat{\alpha}$ & 0.183056 & 0.217668 & 0.502692\\
20 & Proposed 2 & $\widehat{\alpha}$ & 0.182812 & 0.219201 & 0.504607\\
20 & ML & $\widehat{\alpha}$ & 0.183249 & 0.217150 & 0.500745\\
20 & Zhao et al. & $\widehat{\alpha}$ & 0.342385 & 0.284307 & 0.708591\\
20 & Nawa--Nadarajah & $\widehat{\alpha}$ & 0.255559 & 0.294804 & 0.724317\\
\cellcolor{gray!10}{20} & \cellcolor{gray!10}{Proposed 1} & \cellcolor{gray!10}{$\widehat{\beta}$} & \cellcolor{gray!10}{0.154651} & \cellcolor{gray!10}{0.211424} & \cellcolor{gray!10}{0.430040}\\
\cellcolor{gray!10}{20} & \cellcolor{gray!10}{Proposed 2} & \cellcolor{gray!10}{$\widehat{\beta}$} & \cellcolor{gray!10}{0.149460} & \cellcolor{gray!10}{0.212082} & \cellcolor{gray!10}{0.429360}\\
\cellcolor{gray!10}{20} & \cellcolor{gray!10}{ML} & \cellcolor{gray!10}{$\widehat{\beta}$} & \cellcolor{gray!10}{0.151779} & \cellcolor{gray!10}{0.211757} & \cellcolor{gray!10}{0.429090}\\
\cellcolor{gray!10}{20} & \cellcolor{gray!10}{Zhao et al.} & \cellcolor{gray!10}{$\widehat{\beta}$} & \cellcolor{gray!10}{0.638517} & \cellcolor{gray!10}{0.483589} & \cellcolor{gray!10}{1.653866}\\
\cellcolor{gray!10}{20} & \cellcolor{gray!10}{Nawa--Nadarajah} & \cellcolor{gray!10}{$\widehat{\beta}$} & \cellcolor{gray!10}{0.217070} & \cellcolor{gray!10}{0.290568} & \cellcolor{gray!10}{0.635391}\\
20 & Proposed 1 & $\widehat{\gamma}$ & 0.135249 & 0.230305 & 0.346751\\
20 & Proposed 2 & $\widehat{\gamma}$ & 0.133441 & 0.229927 & 0.346013\\
20 & ML & $\widehat{\gamma}$ & 0.134333 & 0.229962 & 0.345806\\
20 & Zhao et al. & $\widehat{\gamma}$ & 0.363722 & 0.392576 & 0.846386\\
20 & Nawa--Nadarajah & $\widehat{\gamma}$ & 0.184599 & 0.310128 & 0.502847\\
\cellcolor{gray!10}{50} & \cellcolor{gray!10}{Proposed 1} & \cellcolor{gray!10}{$\widehat{\alpha}$} & \cellcolor{gray!10}{0.050310} & \cellcolor{gray!10}{0.117264} & \cellcolor{gray!10}{0.254743}\\
\cellcolor{gray!10}{50} & \cellcolor{gray!10}{Proposed 2} & \cellcolor{gray!10}{$\widehat{\alpha}$} & \cellcolor{gray!10}{0.052324} & \cellcolor{gray!10}{0.117185} & \cellcolor{gray!10}{0.255945}\\
\cellcolor{gray!10}{50} & \cellcolor{gray!10}{ML} & \cellcolor{gray!10}{$\widehat{\alpha}$} & \cellcolor{gray!10}{0.051081} & \cellcolor{gray!10}{0.116408} & \cellcolor{gray!10}{0.252802}\\
\cellcolor{gray!10}{50} & \cellcolor{gray!10}{Zhao et al.} & \cellcolor{gray!10}{$\widehat{\alpha}$} & \cellcolor{gray!10}{0.118499} & \cellcolor{gray!10}{0.138044} & \cellcolor{gray!10}{0.303087}\\
\cellcolor{gray!10}{50} & \cellcolor{gray!10}{Nawa--Nadarajah} & \cellcolor{gray!10}{$\widehat{\alpha}$} & \cellcolor{gray!10}{0.084444} & \cellcolor{gray!10}{0.159489} & \cellcolor{gray!10}{0.359685}\\
50 & Proposed 1 & $\widehat{\beta}$ & 0.050373 & 0.117359 & 0.225154\\
50 & Proposed 2 & $\widehat{\beta}$ & 0.046967 & 0.116489 & 0.222827\\
50 & ML & $\widehat{\beta}$ & 0.048490 & 0.116590 & 0.222857\\
50 & Zhao et al. & $\widehat{\beta}$ & 0.263738 & 0.233676 & 0.511098\\
50 & Nawa--Nadarajah & $\widehat{\beta}$ & 0.076237 & 0.161318 & 0.316784\\
\cellcolor{gray!10}{50} & \cellcolor{gray!10}{Proposed 1} & \cellcolor{gray!10}{$\widehat{\gamma}$} & \cellcolor{gray!10}{0.037518} & \cellcolor{gray!10}{0.121546} & \cellcolor{gray!10}{0.172831}\\
\cellcolor{gray!10}{50} & \cellcolor{gray!10}{Proposed 2} & \cellcolor{gray!10}{$\widehat{\gamma}$} & \cellcolor{gray!10}{0.037054} & \cellcolor{gray!10}{0.121290} & \cellcolor{gray!10}{0.172042}\\
\cellcolor{gray!10}{50} & \cellcolor{gray!10}{ML} & \cellcolor{gray!10}{$\widehat{\gamma}$} & \cellcolor{gray!10}{0.037170} & \cellcolor{gray!10}{0.120901} & \cellcolor{gray!10}{0.171942}\\
\cellcolor{gray!10}{50} & \cellcolor{gray!10}{Zhao et al.} & \cellcolor{gray!10}{$\widehat{\gamma}$} & \cellcolor{gray!10}{0.135090} & \cellcolor{gray!10}{0.176559} & \cellcolor{gray!10}{0.272566}\\
\cellcolor{gray!10}{50} & \cellcolor{gray!10}{Nawa--Nadarajah} & \cellcolor{gray!10}{$\widehat{\gamma}$} & \cellcolor{gray!10}{0.058158} & \cellcolor{gray!10}{0.164665} & \cellcolor{gray!10}{0.238681}\\
100 & Proposed 1 & $\widehat{\alpha}$ & 0.022715 & 0.082044 & 0.175573\\
100 & Proposed 2 & $\widehat{\alpha}$ & 0.022485 & 0.081859 & 0.174847\\
100 & ML & $\widehat{\alpha}$ & 0.023641 & 0.080263 & 0.171695\\
100 & Zhao et al. & $\widehat{\alpha}$ & 0.068795 & 0.092263 & 0.203373\\
100 & Nawa--Nadarajah & $\widehat{\alpha}$ & 0.029152 & 0.107630 & 0.232769\\
\cellcolor{gray!10}{100} & \cellcolor{gray!10}{Proposed 1} & \cellcolor{gray!10}{$\widehat{\beta}$} & \cellcolor{gray!10}{0.019464} & \cellcolor{gray!10}{0.086850} & \cellcolor{gray!10}{0.163817}\\
\cellcolor{gray!10}{100} & \cellcolor{gray!10}{Proposed 2} & \cellcolor{gray!10}{$\widehat{\beta}$} & \cellcolor{gray!10}{0.019598} & \cellcolor{gray!10}{0.084624} & \cellcolor{gray!10}{0.159224}\\
\cellcolor{gray!10}{100} & \cellcolor{gray!10}{ML} & \cellcolor{gray!10}{$\widehat{\beta}$} & \cellcolor{gray!10}{0.019729} & \cellcolor{gray!10}{0.084603} & \cellcolor{gray!10}{0.159102}\\
\cellcolor{gray!10}{100} & \cellcolor{gray!10}{Zhao et al.} & \cellcolor{gray!10}{$\widehat{\beta}$} & \cellcolor{gray!10}{0.150519} & \cellcolor{gray!10}{0.160661} & \cellcolor{gray!10}{0.326007}\\
\cellcolor{gray!10}{100} & \cellcolor{gray!10}{Nawa--Nadarajah} & \cellcolor{gray!10}{$\widehat{\beta}$} & \cellcolor{gray!10}{0.024699} & \cellcolor{gray!10}{0.110045} & \cellcolor{gray!10}{0.208142}\\
100 & Proposed 1 & $\widehat{\gamma}$ & 0.018940 & 0.088100 & 0.121902\\
100 & Proposed 2 & $\widehat{\gamma}$ & 0.018853 & 0.086040 & 0.119157\\
100 & ML & $\widehat{\gamma}$ & 0.019313 & 0.085790 & 0.118997\\
100 & Zhao et al. & $\widehat{\gamma}$ & 0.079627 & 0.120690 & 0.175980\\
100 & Nawa--Nadarajah & $\widehat{\gamma}$ & 0.022792 & 0.111203 & 0.153922\\
\bottomrule
\end{tabular}
\end{table}

\begin{table}[!ht]
\centering
\vskip -1.5cm
\caption{Monte Carlo summary for scenario $(\alpha,\beta,\gamma)=(3.0,1.0,2.0)$. Reported metrics per sample size $n$ and estimator.}\label{tab:sim2}
\begin{tabular}{rllrrr}
\toprule
$n$ & Estimator & Parameter & AB & MARE & RMSE\\
\midrule
20 & Proposed 1 & $\widehat{\alpha}$ & 0.436795 & 0.240443 & 0.996968\\
20 & Proposed 2 & $\widehat{\alpha}$ & 0.329578 & 0.232517 & 0.955157\\
20 & ML & $\widehat{\alpha}$ & 0.338034 & 0.227814 & 0.951147\\
20 & Zhao et al. & $\widehat{\alpha}$ & 0.838085 & 0.347155 & 2.331154\\
20 & Nawa--Nadarajah & $\widehat{\alpha}$ & 0.481693 & 0.325206 & 1.379224\\
\cellcolor{gray!10}{20} & \cellcolor{gray!10}{Proposed 1} & \cellcolor{gray!10}{$\widehat{\beta}$} & \cellcolor{gray!10}{0.163041} & \cellcolor{gray!10}{0.210408} & \cellcolor{gray!10}{0.302224}\\
\cellcolor{gray!10}{20} & \cellcolor{gray!10}{Proposed 2} & \cellcolor{gray!10}{$\widehat{\beta}$} & \cellcolor{gray!10}{0.090750} & \cellcolor{gray!10}{0.217111} & \cellcolor{gray!10}{0.296095}\\
\cellcolor{gray!10}{20} & \cellcolor{gray!10}{ML} & \cellcolor{gray!10}{$\widehat{\beta}$} & \cellcolor{gray!10}{0.094818} & \cellcolor{gray!10}{0.215901} & \cellcolor{gray!10}{0.296042}\\
\cellcolor{gray!10}{20} & \cellcolor{gray!10}{Zhao et al.} & \cellcolor{gray!10}{$\widehat{\beta}$} & \cellcolor{gray!10}{0.521916} & \cellcolor{gray!10}{0.536654} & \cellcolor{gray!10}{1.495185}\\
\cellcolor{gray!10}{20} & \cellcolor{gray!10}{Nawa--Nadarajah} & \cellcolor{gray!10}{$\widehat{\beta}$} & \cellcolor{gray!10}{0.136079} & \cellcolor{gray!10}{0.286414} & \cellcolor{gray!10}{0.422009}\\
20 & Proposed 1 & $\widehat{\gamma}$ & 0.327198 & 0.245341 & 0.670004\\
20 & Proposed 2 & $\widehat{\gamma}$ & 0.236789 & 0.236096 & 0.641065\\
20 & ML & $\widehat{\gamma}$ & 0.243164 & 0.234699 & 0.642509\\
20 & Zhao et al. & $\widehat{\gamma}$ & 0.710237 & 0.400387 & 1.911178\\
20 & Nawa--Nadarajah & $\widehat{\gamma}$ & 0.334952 & 0.325573 & 0.905578\\
\cellcolor{gray!10}{50} & \cellcolor{gray!10}{Proposed 1} & \cellcolor{gray!10}{$\widehat{\alpha}$} & \cellcolor{gray!10}{0.194844} & \cellcolor{gray!10}{0.135794} & \cellcolor{gray!10}{0.531719}\\
\cellcolor{gray!10}{50} & \cellcolor{gray!10}{Proposed 2} & \cellcolor{gray!10}{$\widehat{\alpha}$} & \cellcolor{gray!10}{0.110297} & \cellcolor{gray!10}{0.125328} & \cellcolor{gray!10}{0.491467}\\
\cellcolor{gray!10}{50} & \cellcolor{gray!10}{ML} & \cellcolor{gray!10}{$\widehat{\alpha}$} & \cellcolor{gray!10}{0.114894} & \cellcolor{gray!10}{0.122168} & \cellcolor{gray!10}{0.473402}\\
\cellcolor{gray!10}{50} & \cellcolor{gray!10}{Zhao et al.} & \cellcolor{gray!10}{$\widehat{\alpha}$} & \cellcolor{gray!10}{0.452191} & \cellcolor{gray!10}{0.187756} & \cellcolor{gray!10}{1.360337}\\
\cellcolor{gray!10}{50} & \cellcolor{gray!10}{Nawa--Nadarajah} & \cellcolor{gray!10}{$\widehat{\alpha}$} & \cellcolor{gray!10}{0.159717} & \cellcolor{gray!10}{0.178343} & \cellcolor{gray!10}{0.736247}\\
50 & Proposed 1 & $\widehat{\beta}$ & 0.094810 & 0.127407 & 0.166462\\
50 & Proposed 2 & $\widehat{\beta}$ & 0.031062 & 0.115799 & 0.149459\\
50 & ML & $\widehat{\beta}$ & 0.032599 & 0.115707 & 0.149220\\
50 & Zhao et al. & $\widehat{\beta}$ & 0.346529 & 0.352497 & 1.406319\\
50 & Nawa--Nadarajah & $\widehat{\beta}$ & 0.043033 & 0.151672 & 0.201890\\
\cellcolor{gray!10}{50} & \cellcolor{gray!10}{Proposed 1} & \cellcolor{gray!10}{$\widehat{\gamma}$} & \cellcolor{gray!10}{0.157027} & \cellcolor{gray!10}{0.139096} & \cellcolor{gray!10}{0.362973}\\
\cellcolor{gray!10}{50} & \cellcolor{gray!10}{Proposed 2} & \cellcolor{gray!10}{$\widehat{\gamma}$} & \cellcolor{gray!10}{0.082235} & \cellcolor{gray!10}{0.127318} & \cellcolor{gray!10}{0.329613}\\
\cellcolor{gray!10}{50} & \cellcolor{gray!10}{ML} & \cellcolor{gray!10}{$\widehat{\gamma}$} & \cellcolor{gray!10}{0.085241} & \cellcolor{gray!10}{0.125948} & \cellcolor{gray!10}{0.324377}\\
\cellcolor{gray!10}{50} & \cellcolor{gray!10}{Zhao et al.} & \cellcolor{gray!10}{$\widehat{\gamma}$} & \cellcolor{gray!10}{0.412028} & \cellcolor{gray!10}{0.226642} & \cellcolor{gray!10}{1.378785}\\
\cellcolor{gray!10}{50} & \cellcolor{gray!10}{Nawa--Nadarajah} & \cellcolor{gray!10}{$\widehat{\gamma}$} & \cellcolor{gray!10}{0.113488} & \cellcolor{gray!10}{0.177293} & \cellcolor{gray!10}{0.480282}\\
100 & Proposed 1 & $\widehat{\alpha}$ & 0.120027 & 0.094627 & 0.362029\\
100 & Proposed 2 & $\widehat{\alpha}$ & 0.055339 & 0.088620 & 0.337063\\
100 & ML & $\widehat{\alpha}$ & 0.052551 & 0.085091 & 0.323044\\
100 & Zhao et al. & $\widehat{\alpha}$ & 0.851859 & 0.310114 & 17.159372\\
100 & Nawa--Nadarajah & $\widehat{\alpha}$ & 0.086258 & 0.121339 & 0.477052\\
\cellcolor{gray!10}{100} & \cellcolor{gray!10}{Proposed 1} & \cellcolor{gray!10}{$\widehat{\beta}$} & \cellcolor{gray!10}{0.077360} & \cellcolor{gray!10}{0.104929} & \cellcolor{gray!10}{0.131812}\\
\cellcolor{gray!10}{100} & \cellcolor{gray!10}{Proposed 2} & \cellcolor{gray!10}{$\widehat{\beta}$} & \cellcolor{gray!10}{0.019921} & \cellcolor{gray!10}{0.084767} & \cellcolor{gray!10}{0.106642}\\
\cellcolor{gray!10}{100} & \cellcolor{gray!10}{ML} & \cellcolor{gray!10}{$\widehat{\beta}$} & \cellcolor{gray!10}{0.020965} & \cellcolor{gray!10}{0.084078} & \cellcolor{gray!10}{0.105568}\\
\cellcolor{gray!10}{100} & \cellcolor{gray!10}{Zhao et al.} & \cellcolor{gray!10}{$\widehat{\beta}$} & \cellcolor{gray!10}{0.732356} & \cellcolor{gray!10}{0.736254} & \cellcolor{gray!10}{14.728715}\\
\cellcolor{gray!10}{100} & \cellcolor{gray!10}{Nawa--Nadarajah} & \cellcolor{gray!10}{$\widehat{\beta}$} & \cellcolor{gray!10}{0.028684} & \cellcolor{gray!10}{0.105818} & \cellcolor{gray!10}{0.136233}\\
100 & Proposed 1 & $\widehat{\gamma}$ & 0.103085 & 0.096609 & 0.247774\\
100 & Proposed 2 & $\widehat{\gamma}$ & 0.041939 & 0.088249 & 0.223093\\
100 & ML & $\widehat{\gamma}$ & 0.041094 & 0.086866 & 0.219275\\
100 & Zhao et al. & $\widehat{\gamma}$ & 0.791494 & 0.409716 & 15.751735\\
100 & Nawa--Nadarajah & $\widehat{\gamma}$ & 0.061678 & 0.119011 & 0.307962\\
\bottomrule
\end{tabular}
\end{table}

\begin{table}[!ht]
\centering
\vskip -1.5cm
\caption{Monte Carlo summary for scenario $(\alpha,\beta,\gamma)=(2.5,4.0,0.6)$. Reported metrics per sample size $n$ and estimator.}\label{tab:sim3}
\begin{tabular}{rllrrr}
\toprule
$n$ & Estimator & Parameter & AB & MARE & RMSE\\
\midrule
20 & Proposed 1 & $\widehat{\alpha}$ & 0.250687 & 0.233236 & 0.765545\\
20 & Proposed 2 & $\widehat{\alpha}$ & 0.289691 & 0.227704 & 0.761350\\
20 & ML & $\widehat{\alpha}$ & 0.290180 & 0.228482 & 0.762401\\
20 & Zhao et al. & $\widehat{\alpha}$ & 0.337064 & 0.242190 & 0.809241\\
20 & Nawa--Nadarajah & $\widehat{\alpha}$ & 0.397047 & 0.295018 & 1.035896\\
\cellcolor{gray!10}{20} & \cellcolor{gray!10}{Proposed 1} & \cellcolor{gray!10}{$\widehat{\beta}$} & \cellcolor{gray!10}{0.398158} & \cellcolor{gray!10}{0.236924} & \cellcolor{gray!10}{1.266435}\\
\cellcolor{gray!10}{20} & \cellcolor{gray!10}{Proposed 2} & \cellcolor{gray!10}{$\widehat{\beta}$} & \cellcolor{gray!10}{0.475394} & \cellcolor{gray!10}{0.226040} & \cellcolor{gray!10}{1.218688}\\
\cellcolor{gray!10}{20} & \cellcolor{gray!10}{ML} & \cellcolor{gray!10}{$\widehat{\beta}$} & \cellcolor{gray!10}{0.475831} & \cellcolor{gray!10}{0.225807} & \cellcolor{gray!10}{1.217572}\\
\cellcolor{gray!10}{20} & \cellcolor{gray!10}{Zhao et al.} & \cellcolor{gray!10}{$\widehat{\beta}$} & \cellcolor{gray!10}{0.626676} & \cellcolor{gray!10}{0.259490} & \cellcolor{gray!10}{1.419108}\\
\cellcolor{gray!10}{20} & \cellcolor{gray!10}{Nawa--Nadarajah} & \cellcolor{gray!10}{$\widehat{\beta}$} & \cellcolor{gray!10}{0.671978} & \cellcolor{gray!10}{0.308901} & \cellcolor{gray!10}{1.769251}\\
20 & Proposed 1 & $\widehat{\gamma}$ & 0.065285 & 0.236705 & 0.189586\\
20 & Proposed 2 & $\widehat{\gamma}$ & 0.076473 & 0.233180 & 0.187629\\
20 & ML & $\widehat{\gamma}$ & 0.076551 & 0.233100 & 0.187598\\
20 & Zhao et al. & $\widehat{\gamma}$ & 0.094977 & 0.255970 & 0.207848\\
20 & Nawa--Nadarajah & $\widehat{\gamma}$ & 0.105173 & 0.311380 & 0.265056\\
\cellcolor{gray!10}{50} & \cellcolor{gray!10}{Proposed 1} & \cellcolor{gray!10}{$\widehat{\alpha}$} & \cellcolor{gray!10}{0.067882} & \cellcolor{gray!10}{0.118793} & \cellcolor{gray!10}{0.387688}\\
\cellcolor{gray!10}{50} & \cellcolor{gray!10}{Proposed 2} & \cellcolor{gray!10}{$\widehat{\alpha}$} & \cellcolor{gray!10}{0.086545} & \cellcolor{gray!10}{0.118547} & \cellcolor{gray!10}{0.386098}\\
\cellcolor{gray!10}{50} & \cellcolor{gray!10}{ML} & \cellcolor{gray!10}{$\widehat{\alpha}$} & \cellcolor{gray!10}{0.086682} & \cellcolor{gray!10}{0.117798} & \cellcolor{gray!10}{0.384936}\\
\cellcolor{gray!10}{50} & \cellcolor{gray!10}{Zhao et al.} & \cellcolor{gray!10}{$\widehat{\alpha}$} & \cellcolor{gray!10}{0.105582} & \cellcolor{gray!10}{0.122635} & \cellcolor{gray!10}{0.401368}\\
\cellcolor{gray!10}{50} & \cellcolor{gray!10}{Nawa--Nadarajah} & \cellcolor{gray!10}{$\widehat{\alpha}$} & \cellcolor{gray!10}{0.127264} & \cellcolor{gray!10}{0.166842} & \cellcolor{gray!10}{0.533607}\\
50 & Proposed 1 & $\widehat{\beta}$ & 0.111188 & 0.133093 & 0.689373\\
50 & Proposed 2 & $\widehat{\beta}$ & 0.142613 & 0.123023 & 0.642274\\
50 & ML & $\widehat{\beta}$ & 0.142523 & 0.123136 & 0.642212\\
50 & Zhao et al. & $\widehat{\beta}$ & 0.204481 & 0.146537 & 0.754890\\
50 & Nawa--Nadarajah & $\widehat{\beta}$ & 0.212947 & 0.177115 & 0.909631\\
\cellcolor{gray!10}{50} & \cellcolor{gray!10}{Proposed 1} & \cellcolor{gray!10}{$\widehat{\gamma}$} & \cellcolor{gray!10}{0.017328} & \cellcolor{gray!10}{0.126968} & \cellcolor{gray!10}{0.099172}\\
\cellcolor{gray!10}{50} & \cellcolor{gray!10}{Proposed 2} & \cellcolor{gray!10}{$\widehat{\gamma}$} & \cellcolor{gray!10}{0.021998} & \cellcolor{gray!10}{0.122744} & \cellcolor{gray!10}{0.096406}\\
\cellcolor{gray!10}{50} & \cellcolor{gray!10}{ML} & \cellcolor{gray!10}{$\widehat{\gamma}$} & \cellcolor{gray!10}{0.021991} & \cellcolor{gray!10}{0.122419} & \cellcolor{gray!10}{0.096277}\\
\cellcolor{gray!10}{50} & \cellcolor{gray!10}{Zhao et al.} & \cellcolor{gray!10}{$\widehat{\gamma}$} & \cellcolor{gray!10}{0.029543} & \cellcolor{gray!10}{0.137176} & \cellcolor{gray!10}{0.106937}\\
\cellcolor{gray!10}{50} & \cellcolor{gray!10}{Nawa--Nadarajah} & \cellcolor{gray!10}{$\widehat{\gamma}$} & \cellcolor{gray!10}{0.032366} & \cellcolor{gray!10}{0.174374} & \cellcolor{gray!10}{0.135257}\\
100 & Proposed 1 & $\widehat{\alpha}$ & 0.030249 & 0.083933 & 0.266960\\
100 & Proposed 2 & $\widehat{\alpha}$ & 0.044811 & 0.080678 & 0.258593\\
100 & ML & $\widehat{\alpha}$ & 0.043453 & 0.080654 & 0.258356\\
100 & Zhao et al. & $\widehat{\alpha}$ & 0.053092 & 0.085354 & 0.271687\\
100 & Nawa--Nadarajah & $\widehat{\alpha}$ & 0.075269 & 0.107486 & 0.344685\\
\cellcolor{gray!10}{100} & \cellcolor{gray!10}{Proposed 1} & \cellcolor{gray!10}{$\widehat{\beta}$} & \cellcolor{gray!10}{0.050882} & \cellcolor{gray!10}{0.090090} & \cellcolor{gray!10}{0.461458}\\
\cellcolor{gray!10}{100} & \cellcolor{gray!10}{Proposed 2} & \cellcolor{gray!10}{$\widehat{\beta}$} & \cellcolor{gray!10}{0.069971} & \cellcolor{gray!10}{0.080188} & \cellcolor{gray!10}{0.416516}\\
\cellcolor{gray!10}{100} & \cellcolor{gray!10}{ML} & \cellcolor{gray!10}{$\widehat{\beta}$} & \cellcolor{gray!10}{0.069807} & \cellcolor{gray!10}{0.079919} & \cellcolor{gray!10}{0.415263}\\
\cellcolor{gray!10}{100} & \cellcolor{gray!10}{Zhao et al.} & \cellcolor{gray!10}{$\widehat{\beta}$} & \cellcolor{gray!10}{0.111640} & \cellcolor{gray!10}{0.095418} & \cellcolor{gray!10}{0.489838}\\
\cellcolor{gray!10}{100} & \cellcolor{gray!10}{Nawa--Nadarajah} & \cellcolor{gray!10}{$\widehat{\beta}$} & \cellcolor{gray!10}{0.123602} & \cellcolor{gray!10}{0.109743} & \cellcolor{gray!10}{0.572774}\\
100 & Proposed 1 & $\widehat{\gamma}$ & 0.008790 & 0.088772 & 0.066827\\
100 & Proposed 2 & $\widehat{\gamma}$ & 0.011926 & 0.083482 & 0.063222\\
100 & ML & $\widehat{\gamma}$ & 0.011786 & 0.083321 & 0.063143\\
100 & Zhao et al. & $\widehat{\gamma}$ & 0.016584 & 0.092349 & 0.070332\\
100 & Nawa--Nadarajah & $\widehat{\gamma}$ & 0.019685 & 0.111548 & 0.085295\\
\bottomrule
\end{tabular}
\end{table}

\begin{table}[!ht]
\centering
\vskip -1.5cm
\caption{Monte Carlo summary for scenario $(\alpha,\beta,\gamma)=(1.2,3.5,1.5)$. Reported metrics per sample size $n$ and estimator.}\label{tab:sim4}
\begin{tabular}{rllrrr}
\toprule
$n$ & Estimator & Parameter & AB & MARE & RMSE\\
\midrule
20 & Proposed 1 & $\widehat{\alpha}$ & 0.119067 & 0.214797 & 0.349245\\
20 & Proposed 2 & $\widehat{\alpha}$ & 0.117049 & 0.216144 & 0.349908\\
20 & ML & $\widehat{\alpha}$ & 0.117473 & 0.215357 & 0.349402\\
20 & Zhao et al. & $\widehat{\alpha}$ & 0.449696 & 0.501067 & 2.319907\\
20 & Nawa--Nadarajah & $\widehat{\alpha}$ & 0.177399 & 0.299355 & 0.507094\\
\cellcolor{gray!10}{20} & \cellcolor{gray!10}{Proposed 1} & \cellcolor{gray!10}{$\widehat{\beta}$} & \cellcolor{gray!10}{0.430784} & \cellcolor{gray!10}{0.237346} & \cellcolor{gray!10}{1.101374}\\
\cellcolor{gray!10}{20} & \cellcolor{gray!10}{Proposed 2} & \cellcolor{gray!10}{$\widehat{\beta}$} & \cellcolor{gray!10}{0.418819} & \cellcolor{gray!10}{0.233488} & \cellcolor{gray!10}{1.090244}\\
\cellcolor{gray!10}{20} & \cellcolor{gray!10}{ML} & \cellcolor{gray!10}{$\widehat{\beta}$} & \cellcolor{gray!10}{0.420402} & \cellcolor{gray!10}{0.233675} & \cellcolor{gray!10}{1.091239}\\
\cellcolor{gray!10}{20} & \cellcolor{gray!10}{Zhao et al.} & \cellcolor{gray!10}{$\widehat{\beta}$} & \cellcolor{gray!10}{3.225042} & \cellcolor{gray!10}{1.231301} & \cellcolor{gray!10}{18.100261}\\
\cellcolor{gray!10}{20} & \cellcolor{gray!10}{Nawa--Nadarajah} & \cellcolor{gray!10}{$\widehat{\beta}$} & \cellcolor{gray!10}{0.640215} & \cellcolor{gray!10}{0.341021} & \cellcolor{gray!10}{1.704572}\\
20 & Proposed 1 & $\widehat{\gamma}$ & 0.191857 & 0.236712 & 0.478226\\
20 & Proposed 2 & $\widehat{\gamma}$ & 0.187203 & 0.234431 & 0.474773\\
20 & ML & $\widehat{\gamma}$ & 0.187831 & 0.234604 & 0.475060\\
20 & Zhao et al. & $\widehat{\gamma}$ & 1.182059 & 1.034751 & 6.971467\\
20 & Nawa--Nadarajah & $\widehat{\gamma}$ & 0.278542 & 0.339522 & 0.721157\\
\cellcolor{gray!10}{50} & \cellcolor{gray!10}{Proposed 1} & \cellcolor{gray!10}{$\widehat{\alpha}$} & \cellcolor{gray!10}{0.043915} & \cellcolor{gray!10}{0.117358} & \cellcolor{gray!10}{0.182963}\\
\cellcolor{gray!10}{50} & \cellcolor{gray!10}{Proposed 2} & \cellcolor{gray!10}{$\widehat{\alpha}$} & \cellcolor{gray!10}{0.042412} & \cellcolor{gray!10}{0.117862} & \cellcolor{gray!10}{0.183151}\\
\cellcolor{gray!10}{50} & \cellcolor{gray!10}{ML} & \cellcolor{gray!10}{$\widehat{\alpha}$} & \cellcolor{gray!10}{0.042886} & \cellcolor{gray!10}{0.117236} & \cellcolor{gray!10}{0.182610}\\
\cellcolor{gray!10}{50} & \cellcolor{gray!10}{Zhao et al.} & \cellcolor{gray!10}{$\widehat{\alpha}$} & \cellcolor{gray!10}{0.357235} & \cellcolor{gray!10}{0.377546} & \cellcolor{gray!10}{3.368394}\\
\cellcolor{gray!10}{50} & \cellcolor{gray!10}{Nawa--Nadarajah} & \cellcolor{gray!10}{$\widehat{\alpha}$} & \cellcolor{gray!10}{0.059233} & \cellcolor{gray!10}{0.159056} & \cellcolor{gray!10}{0.250299}\\
50 & Proposed 1 & $\widehat{\beta}$ & 0.144332 & 0.123246 & 0.574161\\
50 & Proposed 2 & $\widehat{\beta}$ & 0.136788 & 0.121918 & 0.568118\\
50 & ML & $\widehat{\beta}$ & 0.136784 & 0.122255 & 0.569114\\
50 & Zhao et al. & $\widehat{\beta}$ & 3.612657 & 1.244094 & 32.666352\\
50 & Nawa--Nadarajah & $\widehat{\beta}$ & 0.201250 & 0.184606 & 0.838399\\
\cellcolor{gray!10}{50} & \cellcolor{gray!10}{Proposed 1} & \cellcolor{gray!10}{$\widehat{\gamma}$} & \cellcolor{gray!10}{0.062762} & \cellcolor{gray!10}{0.126885} & \cellcolor{gray!10}{0.248284}\\
\cellcolor{gray!10}{50} & \cellcolor{gray!10}{Proposed 2} & \cellcolor{gray!10}{$\widehat{\gamma}$} & \cellcolor{gray!10}{0.059792} & \cellcolor{gray!10}{0.125408} & \cellcolor{gray!10}{0.246486}\\
\cellcolor{gray!10}{50} & \cellcolor{gray!10}{ML} & \cellcolor{gray!10}{$\widehat{\gamma}$} & \cellcolor{gray!10}{0.059954} & \cellcolor{gray!10}{0.125651} & \cellcolor{gray!10}{0.246803}\\
\cellcolor{gray!10}{50} & \cellcolor{gray!10}{Zhao et al.} & \cellcolor{gray!10}{$\widehat{\gamma}$} & \cellcolor{gray!10}{1.282991} & \cellcolor{gray!10}{1.027041} & \cellcolor{gray!10}{11.462926}\\
\cellcolor{gray!10}{50} & \cellcolor{gray!10}{Nawa--Nadarajah} & \cellcolor{gray!10}{$\widehat{\gamma}$} & \cellcolor{gray!10}{0.085593} & \cellcolor{gray!10}{0.180704} & \cellcolor{gray!10}{0.351756}\\
100 & Proposed 1 & $\widehat{\alpha}$ & 0.020199 & 0.083865 & 0.127597\\
100 & Proposed 2 & $\widehat{\alpha}$ & 0.019852 & 0.083863 & 0.128038\\
100 & ML & $\widehat{\alpha}$ & 0.019696 & 0.083636 & 0.127492\\
100 & Zhao et al. & $\widehat{\alpha}$ & 0.143533 & 0.181719 & 1.717795\\
100 & Nawa--Nadarajah & $\widehat{\alpha}$ & 0.028614 & 0.106044 & 0.165524\\
\cellcolor{gray!10}{100} & \cellcolor{gray!10}{Proposed 1} & \cellcolor{gray!10}{$\widehat{\beta}$} & \cellcolor{gray!10}{0.062748} & \cellcolor{gray!10}{0.088682} & \cellcolor{gray!10}{0.392095}\\
\cellcolor{gray!10}{100} & \cellcolor{gray!10}{Proposed 2} & \cellcolor{gray!10}{$\widehat{\beta}$} & \cellcolor{gray!10}{0.058356} & \cellcolor{gray!10}{0.087157} & \cellcolor{gray!10}{0.385674}\\
\cellcolor{gray!10}{100} & \cellcolor{gray!10}{ML} & \cellcolor{gray!10}{$\widehat{\beta}$} & \cellcolor{gray!10}{0.058340} & \cellcolor{gray!10}{0.086932} & \cellcolor{gray!10}{0.384621}\\
\cellcolor{gray!10}{100} & \cellcolor{gray!10}{Zhao et al.} & \cellcolor{gray!10}{$\widehat{\beta}$} & \cellcolor{gray!10}{2.872961} & \cellcolor{gray!10}{0.972697} & \cellcolor{gray!10}{63.809762}\\
\cellcolor{gray!10}{100} & \cellcolor{gray!10}{Nawa--Nadarajah} & \cellcolor{gray!10}{$\widehat{\beta}$} & \cellcolor{gray!10}{0.087342} & \cellcolor{gray!10}{0.119698} & \cellcolor{gray!10}{0.527678}\\
100 & Proposed 1 & $\widehat{\gamma}$ & 0.030195 & 0.089429 & 0.169218\\
100 & Proposed 2 & $\widehat{\gamma}$ & 0.028633 & 0.088108 & 0.167095\\
100 & ML & $\widehat{\gamma}$ & 0.028595 & 0.087976 & 0.166900\\
100 & Zhao et al. & $\widehat{\gamma}$ & 0.988280 & 0.780595 & 21.468651\\
100 & Nawa--Nadarajah & $\widehat{\gamma}$ & 0.040415 & 0.115380 & 0.221184\\
\bottomrule
\end{tabular}
\end{table}

\clearpage

\section{Application to real data}\label{sec:application}

We illustrate the proposed estimation methods with the classic Los Angeles rainfall series
(1878--1996; annual totals in inches); see \cite{Nawa2023} and Table \ref{tab:rain15}. Let $X_t$ denote the yearly rainfall in
year $t$ and define the two-year total $Y_t=X_t+X_{t+1}$, so that the McKay
construction $Y=X+Z$ holds with $Z=X_{t+1}$ and $X\sim\text{Gamma}(\alpha,\gamma)$,
$Z\sim\text{Gamma}(\beta,\gamma)$. Using overlapping
windows, this yields $n=118$ pairs $(X_t,Y_t)$ with $0<X_t<Y_t$.

\setlength{\tabcolsep}{3pt}
\renewcommand{\arraystretch}{1.1}
\begin{longtable}{*{15}{r}}
\caption{Los Angeles annual rainfall (inches).}\label{tab:rain15}\\
\toprule
\multicolumn{15}{c}{Rain (in)}\\
\midrule
20.86 & 17.41 & 18.65 & 5.53 & 10.74 & 14.14 & 40.29 & 10.53 & 16.72 & 16.02 & 20.82 & 33.26 & 12.69 & 12.84 & 18.72 \\
21.96 & 7.51 & 12.55 & 11.80 & 14.28 & 4.83 & 8.69 & 11.30 & 11.96 & 13.12 & 14.77 & 11.88 & 19.19 & 21.46 & 15.30 \\
13.74 & 23.92 & 4.89 & 17.85 & 9.78 & 17.17 & 23.21 & 16.67 & 23.29 & 8.45 & 17.49 & 8.82 & 11.18 & 19.85 & 15.27 \\
6.25 & 8.11 & 8.94 & 18.56 & 18.63 & 8.69 & 8.32 & 13.02 & 18.93 & 10.72 & 18.76 & 14.67 & 14.49 & 18.24 & 17.97 \\
27.16 & 12.06 & 20.26 & 31.28 & 7.40 & 22.57 & 17.45 & 12.78 & 16.22 & 4.13 & 7.59 & 10.63 & 7.38 & 14.33 & 24.95 \\
4.08 & 13.69 & 11.89 & 13.62 & 13.24 & 17.49 & 6.23 & 9.57 & 5.83 & 15.37 & 12.31 & 7.89 & 26.81 & 12.91 & 23.66 \\
7.58 & 26.32 & 16.54 & 9.26 & 6.54 & 17.45 & 16.69 & 10.70 & 11.01 & 14.97 & 30.57 & 17.00 & 26.33 & 10.92 & 14.41 \\
34.04 & 8.90 & 8.92 & 18.00 & 9.11 & 11.57 & 4.56 & 6.49 & 15.07 & 22.65 & 23.44 & 8.69 & 24.06 & 17.75 & {} \\
\bottomrule
\end{longtable}

We consider five estimators discussed earlier: (i) Proposed~1 (Section~\ref{new_proposed}), which gives closed–form
$(\widehat\alpha,\widehat\beta,\widehat\gamma)$ for fixed $(r,s)$ and selects
$(r,s)$ via profiling; (ii) Proposed~2 (Section~\ref{new_proposed-1}),
built from the stochastic relationship between $X$ and $Y$ and likewise profiled
over $(r,s)$; (iii) maximum likelihood (ML);  (iv) the closed-form estimator of
\citet{Zhao2022}; and (v) the closed-form
competitor of \citet{Nawa2023}. Table~\ref{tab:la_rain_with_SEs_cvm} reports point estimates with moving-block bootstrap standard errors (SEs in parentheses) and generalized Cramér--von
Mises (CvM) goodness-of-fit $p$-values computed from Rosenblatt-transformed
pairs $(U_1,U_2)$ under a Uniform$(0,1)^2$ bootstrap null ($B=3000$). In this table, log-likelihood values are also presented.

\begin{table}[!ht]
\centering
\caption{Estimates with bootstrap standard errors (between brackets) and generalized CvM goodness-of-fit $p$-values for Los Angeles rainfall data.}
\label{tab:la_rain_with_SEs_cvm}
\renewcommand{\arraystretch}{1.2}
\begin{tabular}{lrrrrrrr}
\toprule
Estimator & $\widehat{\alpha}$ & $\widehat{\beta}$ & $\widehat{\gamma}$ & $\hat r$ & $\hat s$ & $\log L$ & $p$-value (CvM) \\
\midrule
Proposed 1 &
\makecell[r]{4.790209\\\footnotesize(0.439463)} &
\makecell[r]{4.709444\\\footnotesize(0.444353)} &
\makecell[r]{0.3172439\\\footnotesize(0.031560)} &
\makecell[r]{0.1\\\footnotesize(0.591867)} &
\makecell[r]{0.9\\\footnotesize(0.844178)} &
-770.9905 & 0.948 \\
Proposed 2 &
\makecell[r]{4.841863\\\footnotesize(0.444748)} &
\makecell[r]{4.838904\\\footnotesize(0.446210)} &
\makecell[r]{0.3232923\\\footnotesize(0.031986)} &
\makecell[r]{1.2\\\footnotesize(0.579342)} &
\makecell[r]{1.2\\\footnotesize(0.627615)} &
-770.9438 & 0.976 \\
ML &
\makecell[r]{4.814062\\\footnotesize(0.444643)} &
\makecell[r]{4.808138\\\footnotesize(0.445945)} &
\makecell[r]{0.3213364\\\footnotesize(0.031943)} &
-- & -- & -770.9414 & 0.978 \\
Zhao et al. &
\makecell[r]{4.693702\\\footnotesize(0.434130)} &
\makecell[r]{4.643815\\\footnotesize(0.439887)} &
\makecell[r]{0.3118293\\\footnotesize(0.031127)} &
-- & -- & -771.0115 & 0.966 \\
Nawa--Nadarajah &
\makecell[r]{4.834165\\\footnotesize(0.583728)} &
\makecell[r]{4.871643\\\footnotesize(0.586421)} &
\makecell[r]{0.3241285\\\footnotesize(0.042201)} &
-- & -- & -770.9589 & 0.982 \\
\bottomrule
\end{tabular}
\end{table}

From Table~\ref{tab:la_rain_with_SEs_cvm}, we note that all procedures deliver very similar parameter estimates and large CvM
$p$-values ($\ge 0.94$), providing no evidence against the McKay model in this
dataset. ML attains the highest log-likelihood, as expected, but Proposed 2 is essentially indistinguishable and selects $(\hat r,\hat s)=(1.2,1.2)$. Proposed~1 also performs well
with $(\hat r,\hat s)=(0.1,0.9)$.

\section{Conclusion}\label{sec:conclusion}

In this paper, we proposed two closed-form estimation strategies for the parameters of McKay’s bivariate gamma distribution. The first methodology is based on monotone transformations of the likelihood equations, whereas the second approach derives its estimating equations directly from stochastic relationships between gamma random variables. Both appraoches include, as special cases, previously available estimators in the literature. We established the asymptotic properties of the proposed estimators, showing that they are consistent and asymptotically normal. The simulation study results suggested that the proposed estimators achieve finite-sample accuracy comparable to maximum likelihood estimators while avoiding iterative numerical optimization. In addition, they were shown to outperform the closed-form estimators of \cite{Zhao2022} and \cite{Nawa2023}. The empirical study with Los Angeles rainfall has shown that the two proposed closed-form estimators can be good alternatives to ML in practice.

	\paragraph{Acknowledgements}
 This study was financed in part by the
Coordenação de Aperfeiçoamento de Pessoal de Nível Superior - Brasil (CAPES) - Finance Code 001.
	
	\paragraph{Disclosure statement}
	There are no conflicts of interest to disclose.



\end{document}